%% file: NMFpaper.tex
\documentclass[useAMS,usenatbib]{mn2e}

\usepackage{lscape}
\usepackage{graphicx}
\usepackage{amssymb}
\usepackage{amsmath}
\usepackage{epstopdf}
\usepackage{color}
\usepackage{soul}
\usepackage{multirow}
\usepackage{longtable}
\usepackage{deluxetable}
\usepackage{textcomp}
\usepackage[caption=false]{subfig}
\usepackage{float}
\usepackage{appendix}

\newcommand{\uu}{\textmu m}

\usepackage{hyperref}

\title[Learning the Fundamental Components of Galaxies with NMF]
  {Learning the Fundamental MIR Spectral Components of Galaxies with Non-Negative Matrix Factorisation}\author[P.D. Hurley et al.]{P.D.~Hurley,$^1$\thanks{Email: p.d.hurley@sussex.ac.uk} S.~Oliver,$^1$ D.~Farrah,$^2$ V.~Lebouteiller,$^3$ H. W. W. ~Spoon,$^4$\\
$^1$Astronomy Centre, Department of Physics and Astronomy, University of Sussex, Falmer, Brighton BN1 9QH, UK\\
$^2$Virginia Polytechnic Institute \& State University, Department of Physics,MC 0435, 910 Drillfield Drive, Blacksburg, VA 24061\\ 
$^3$Laboratoire AIM, CEA/DSM-CNRS-Universite? Paris Diderot DAPNIA/Service dÕAstrophysique Baöt. 709, CEA-Saclay, F-91191 \\Gif-sur-Yvette Cedex, France\\
$^4$Department of Astronomy and Center for Radiophysics and Space Research, Cornell University, Space Sciences Building, \\Ithaca, NY 14853-6801, USA}

\date{Released 2002 Xxxxx XX}

\pagerange{\pageref{firstpage}--\pageref{lastpage}} \pubyear{2002}

\def\LaTeX{L\kern-.36em\raise.3ex\hbox{a}\kern-.15em
    T\kern-.1667em\lower.7ex\hbox{E}\kern-.125emX}

\graphicspath{{./Average_templates_7/}{./Evo_network/}{./galaxy_fits/}}
\begin{document}

\label{firstpage}
\maketitle

\begin{abstract}
The mid-infrared (MIR) spectra observed with the \textit{Spitzer} Infrared Spectrograph (IRS) provide a valuable dataset for untangling the physical processes and conditions within galaxies. 

This paper presents the first attempt to blindly learn fundamental spectral components of MIR galaxy spectra, using non-negative matrix factorisation (NMF). NMF is a recently developed multivariate technique shown to be successful in blind source separation problems. Unlike the more popular multivariate analysis technique, principal component analysis, NMF imposes the condition that weights and spectral components are non-negative. This more closely resembles the physical process of emission in the mid-infrared, resulting in physically intuitive components. By applying NMF to galaxy spectra in the Cornell Atlas of Spitzer/IRS sources (CASSIS), we find similar components amongst different NMF sets. These similar components include two for AGN emission and one for star formation. The first AGN component is dominated by fine structure emission lines and hot dust, the second by broad silicate emission at 10 and 18 $\mathrm{\mu m}$. The star formation component contains all the PAH features and molecular hydrogen lines. Other components include rising continuums at longer wavelengths, indicative of colder grey-body dust emission. We show an NMF set with seven components can reconstruct the general spectral shape of a wide variety of objects, though struggle to fit the varying strength of emission lines. We also show that the seven components can be used to separate out different types of objects. We model this separation with Gaussian Mixtures modelling and use the result to provide a classification tool.

We also show the NMF components can be used to separate out the emission from AGN and star formation regions and define a new star formation/AGN diagnostic which is consistent with all mid-infrared diagnostics already in use but has the advantage that it can be applied to mid-infrared spectra with low signal to noise or with limited spectral range. The 7 NMF components and code for classification are made public on arxiv and are available at: \url{https://github.com/pdh21/NMF_software/}.
\end{abstract}

\input{Introduction_v2}
\input{CASSIS}

\input{Matrix_factorisation}
\input{Templatesv3}

\input{Conclusions_v2}

\section*{Acknowledgements}
We thank the referee for the useful comments, which have improved the paper. We acknowledge support from the Science and Technology Facilities Council [grant numbers ST/F006977/1, ST/I000976/1]. This work is based on observations made with the Spitzer Space Telescope, which is operated by the Jet Propulsion Laboratory, California Institute of Technology under a contract with NASA. %
\bibliography{sortedbib-05-07-10.bib}
\appendix
\section{Non-linear matrix factorisation techniques}
ICA, PCA and NMF are linear models and cannot efficiently model non-linearities such as dust extinction. Over the last decade, non-linear matrix factorisation techniques have been developed to overcome certain non-linear situations. All of these nonlinear based techniques use kernels to map data with nonlinear structure into a kernel feature space, where the structure becomes linear. Techniques such as PCA or NMF can then be performed in the kernel feature space to recover the structure. These types of techniques are suited to problems where the non-linearity is of parametric form, e.g. points distributed along a circle. Dust extinction is exponential relationship unsuited to this type of technique (Binbin Pan, private communication).

\section{$NMF_{30}$ and Extinction simulation}\label{App:B}
\begin{figure*}
\includegraphics{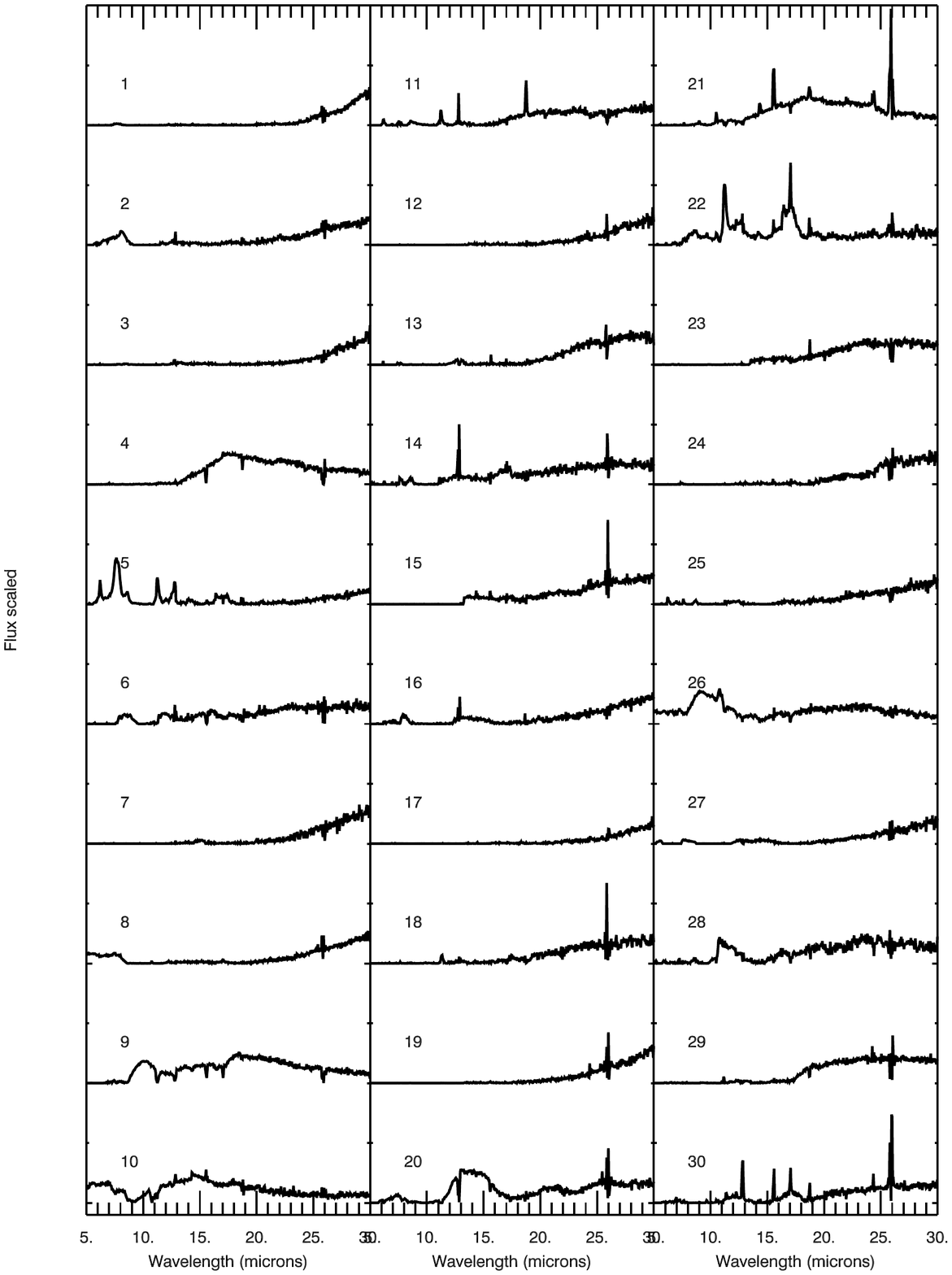}
\caption{The 30 components of NMF set $NMF_{30}$. }
 \label{NMF30}
 \end{figure*}
To explore whether the extinction can cause problems with our NMF analysis, we simulate extinction via equation \ref{eq:simple_ext} described in section \ref{sec:matrix}. 

Our simulation is divided into two parts. The first part assumes galaxy spectra are a linear combination as described in equation \ref{eq:linear}, while the second assumes equation \ref{eq:simple_ext} is valid. To simulate the spectra, we use NMF set $NMF_{5}$ and linearly combine them with weights randomly sampled from a distribution based on those found in the real sample. We do this 500 times to create 500 unique galaxy spectra.

The second part of our simulation involves adding extinction to the simulated spectra as described in equation \ref{eq:simple_ext} in section \ref{sec:matrix}. $\tau(\lambda)$ is defined by the Galactic Centre extinction law of \cite{Chiar:2006}.

We then carry out the NMF algorithm on both the unextincted and extincted spectra. We run the algorithm for $NMF_{5}$-$NMF_{20}$ and use the simplified model selection measures: the Akaike Information Criterion (AIC;\cite{Akaike:1974}) and the Bayesian Information Criterion (BIC; \cite{Schwarz:1978}), defined as follows:

\begin{equation}
AIC \equiv -2\ln L_{max}+2k +\frac{2k(k+1)}{N-k-1}
\end{equation}

\begin{equation}
BIC \equiv -2\ln L_{max} + k\ln N
\end{equation}

\begin{figure}
\includegraphics[width=8.5cm]{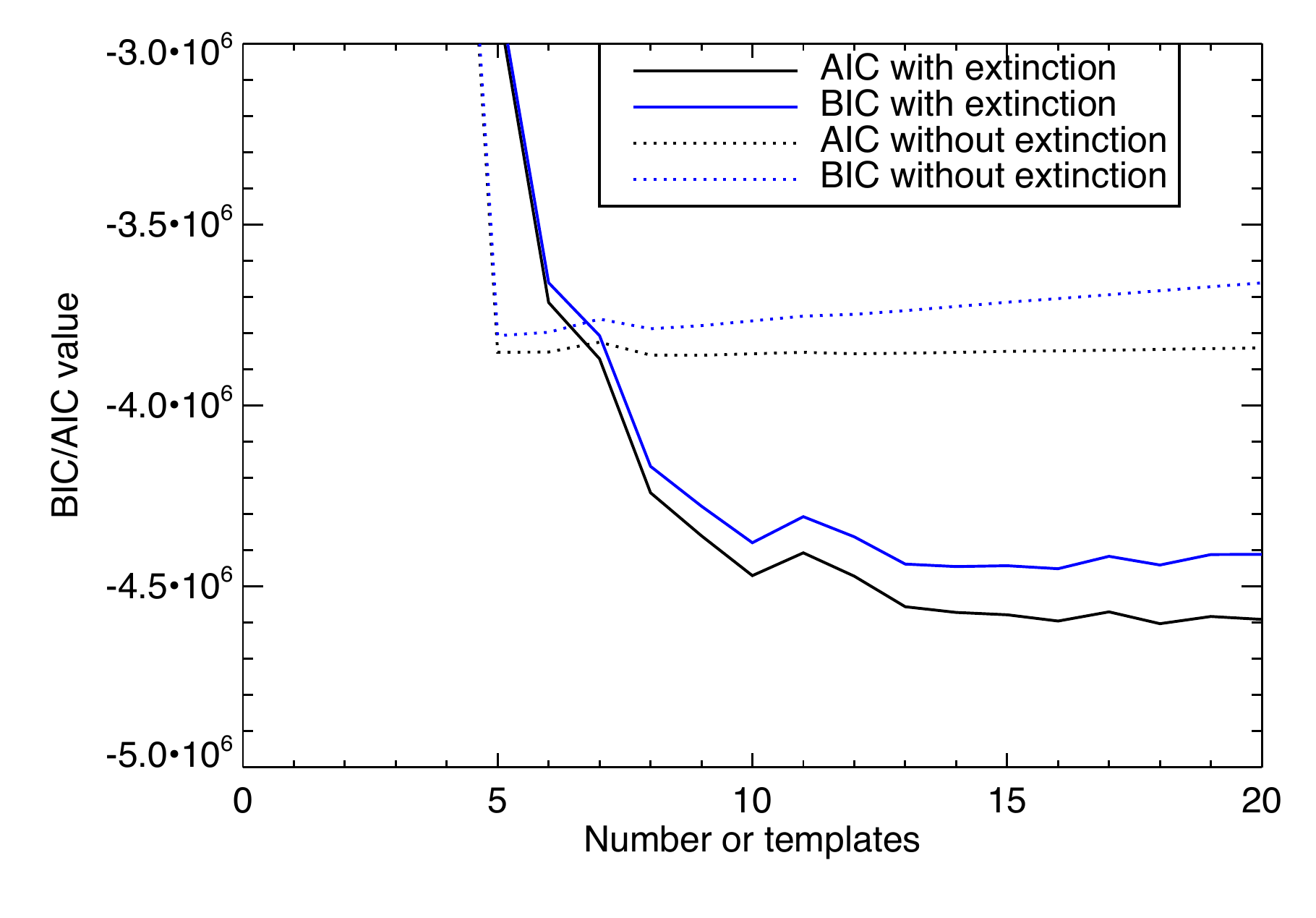}
\caption{The AIC and BIC for the non linear simulations. Both the BIC and AIC for spectra without extinction indicate 5 components as expected. The set with extinction requires around 15-20.}\label{fig:non_lin_sim}
\end{figure}

$L_{max}$ is the maximum likelihood solution, $N$ is the number of datapoints and $k$ is the number of parameters. A minimum value for the AIC and BIC correspond to the optimum model. Figure \ref{fig:non_lin_sim} shows both the BIC and AIC for both sets of simulated spectra. As expected, the BIC and AIC indicate the spectra without extinction can be adequately described by the NMF set with 5 components. For spectra with extinction, the BIC and AIC do not level of until $NMF_{15}$-$NMF_{20}$. This suggests that extinction could be a factor in driving our linear methods to more templates than might be required by underlying physical conditions.

\section{$NMF_{7}$ fits to galaxy spectra}\label{App:C}

\begin{figure*}
\begin{tabular}{c c}
\includegraphics[width=8.5cm]{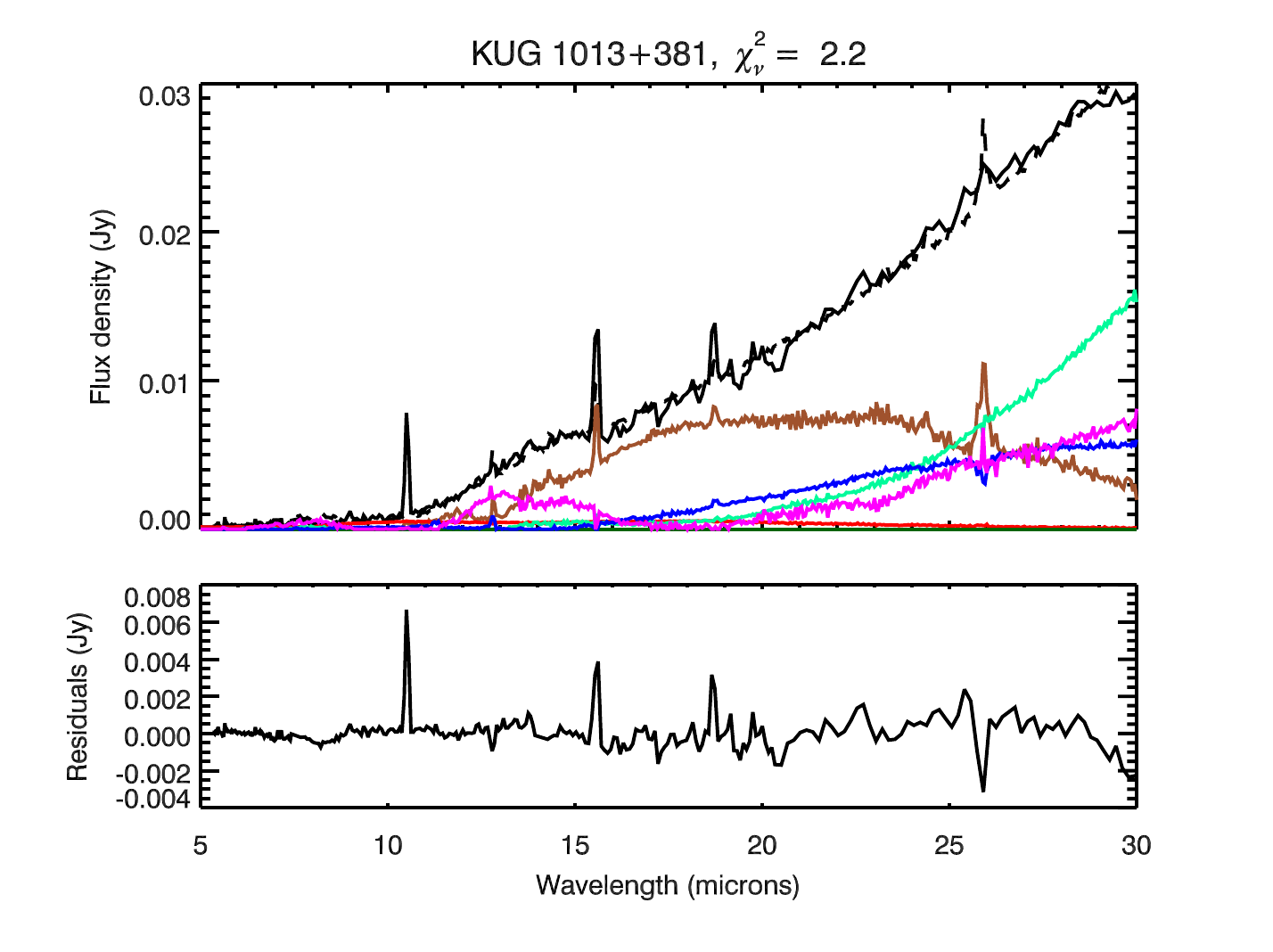} & \includegraphics[width=8.5cm]{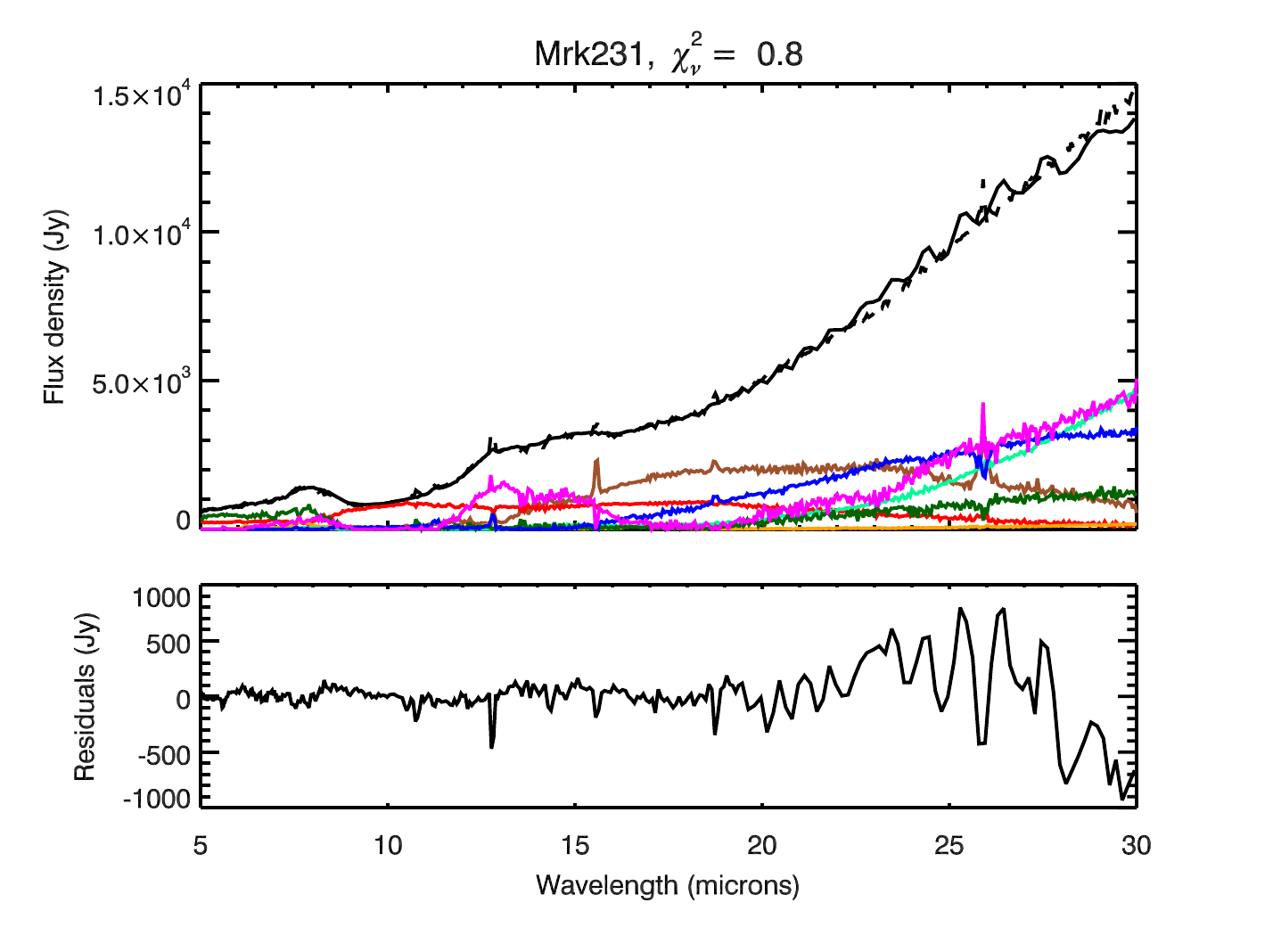} \\
\includegraphics[width=8.5cm]{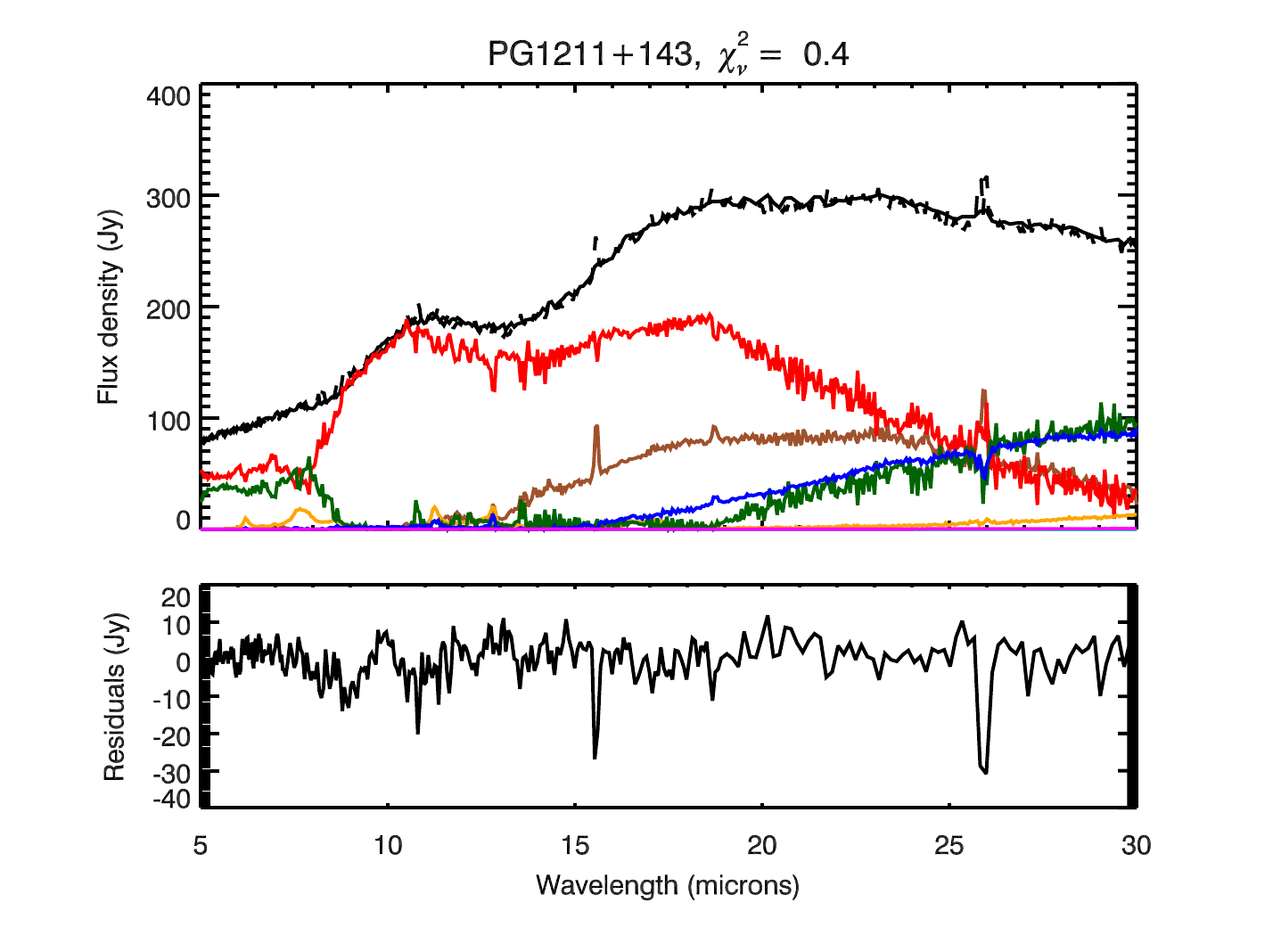} & \includegraphics[width=8.5cm]{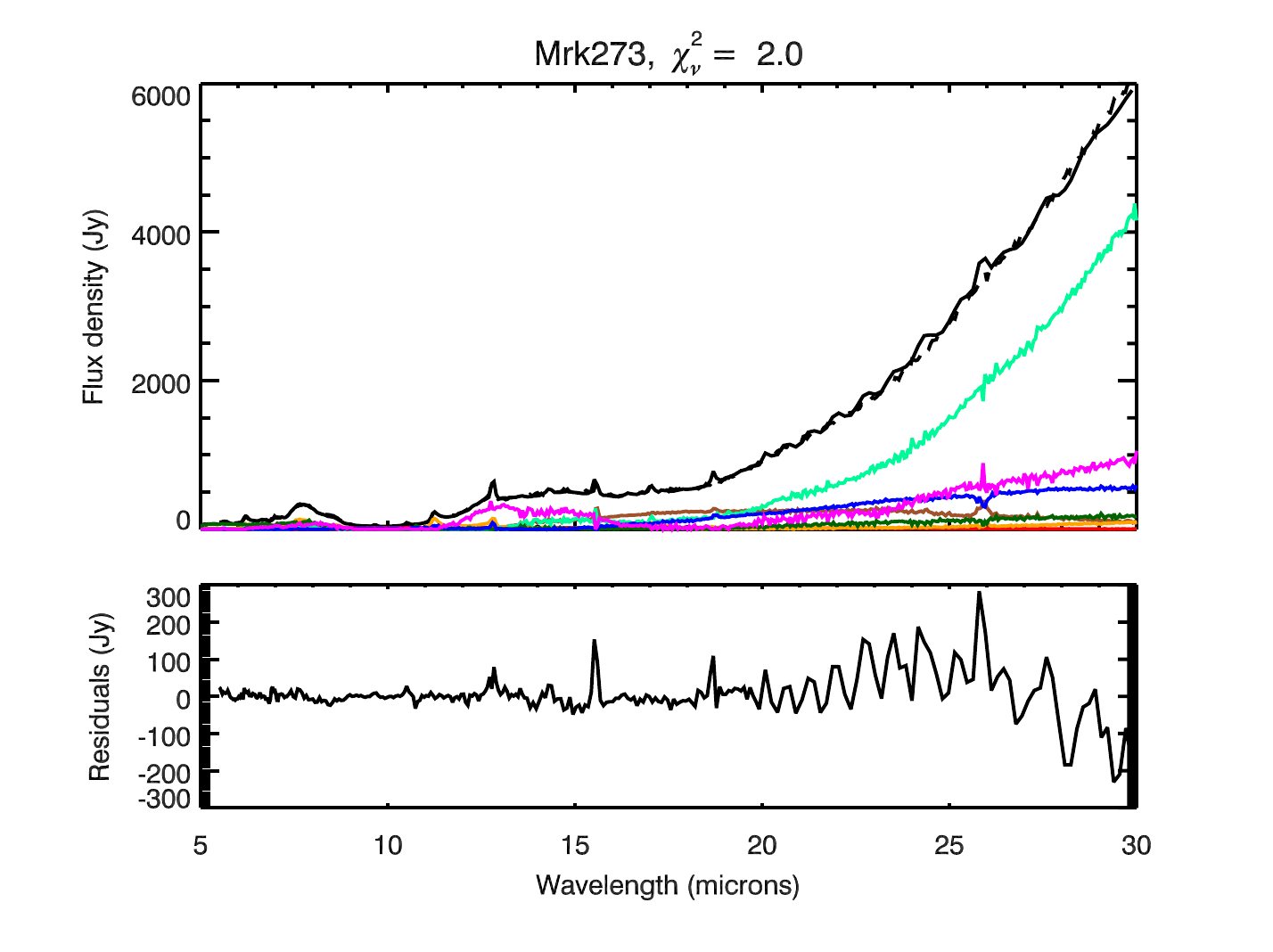} \\
\includegraphics[width=8.5cm]{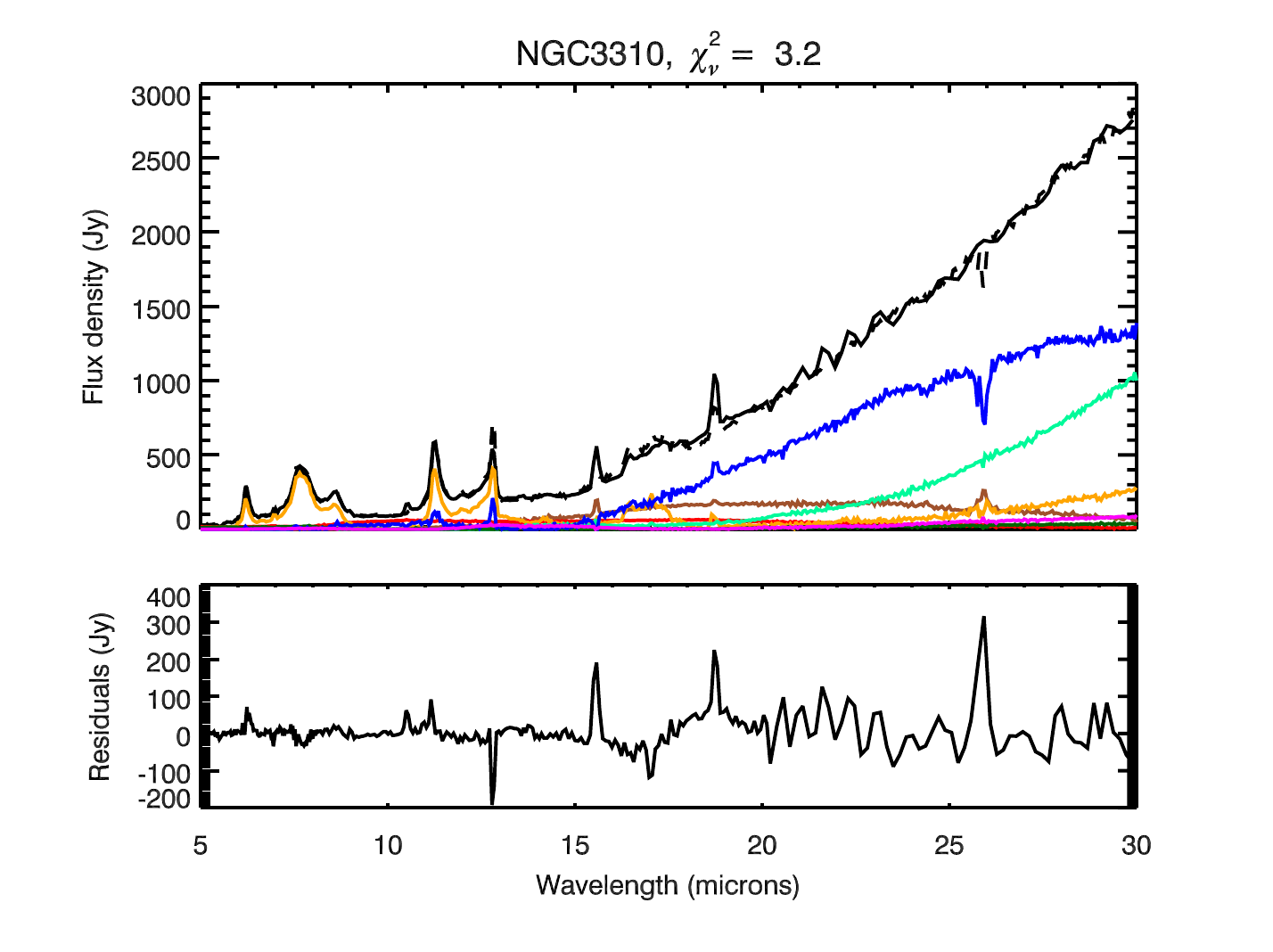} & \includegraphics[width=8.5cm]{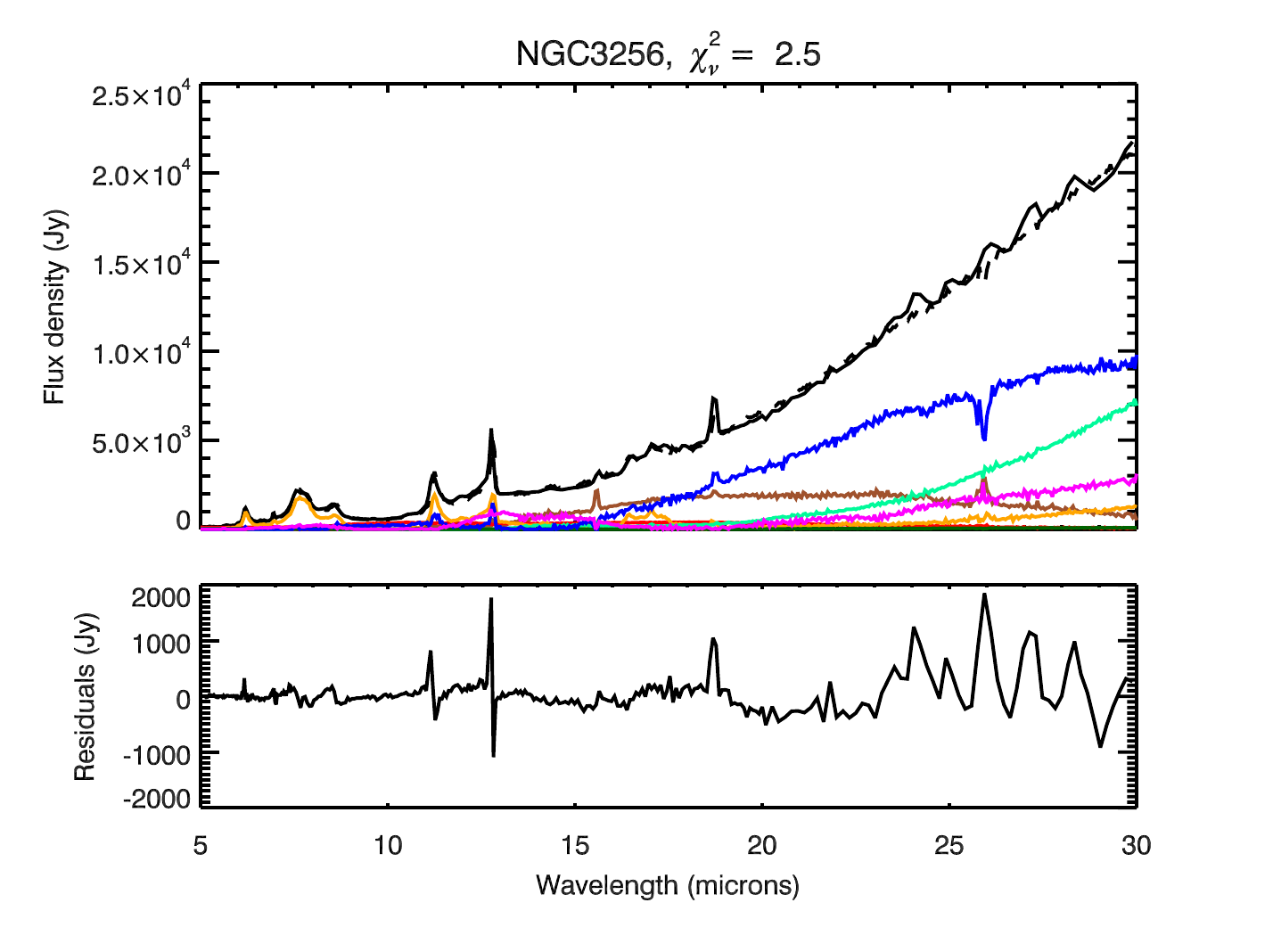} \\
\end{tabular}
\caption{$NMF_{7}$ fits to the Blue Compact Dwarf: KUG 1013+381, Seyfert type 1 galaxies: Markarian 231 and PG1211+143, Seyfert Type 2 galaxy: Markarian 273, and starburst galaxies: NGC3310 and NGC3256. Each spectrum is plotted as a black solid line and the NMF fit as black dashed line. The contribution from each component is also shown, with the same colour coding as in Figure \ref{7_components}. The residuals (data-fit) are plotted below each fit.}\label{example_gal_fits}
\end{figure*}

\begin{figure*}
\begin{tabular}{c c}
\includegraphics[width=8.5cm]{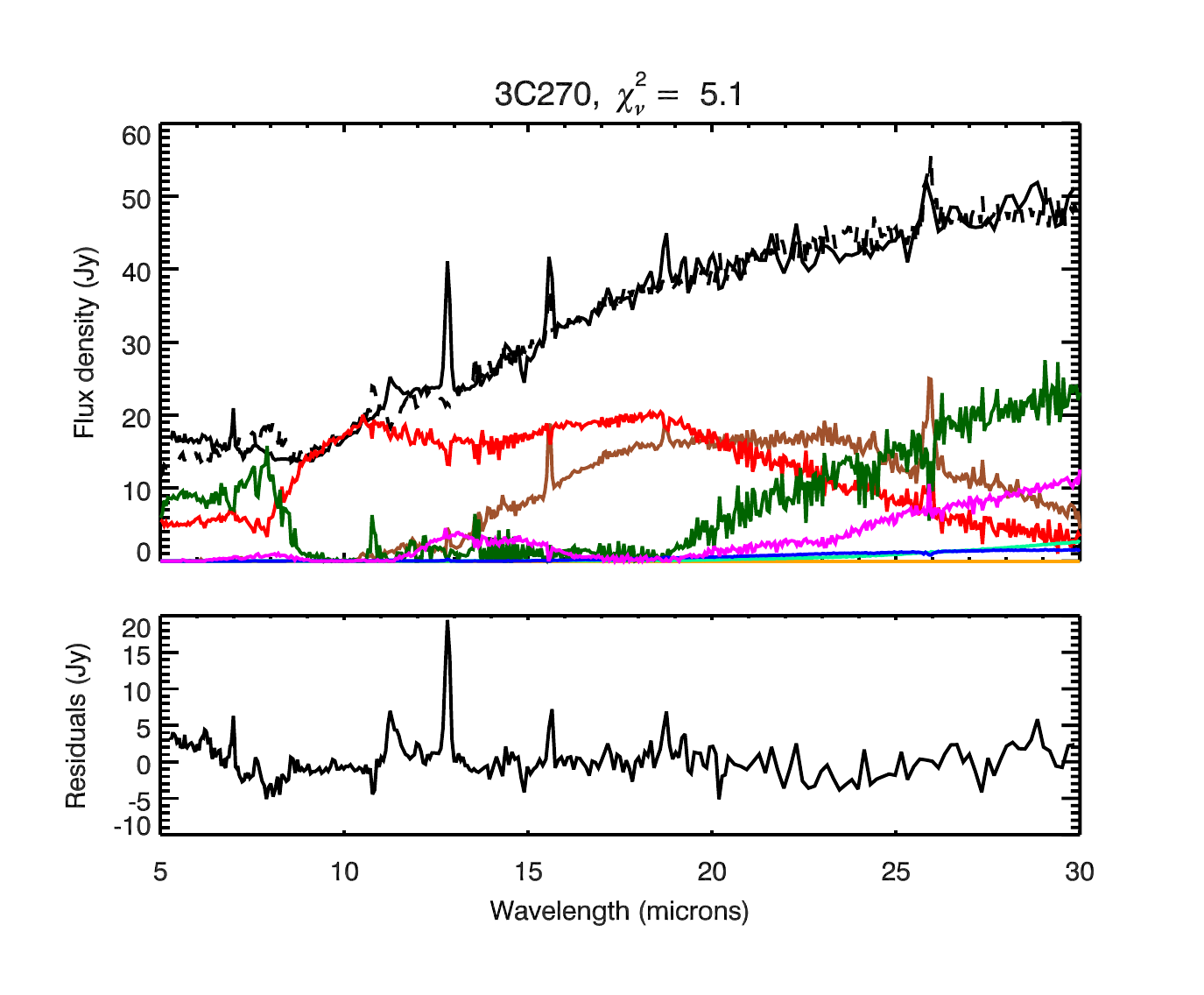} & \includegraphics[width=8.5cm]{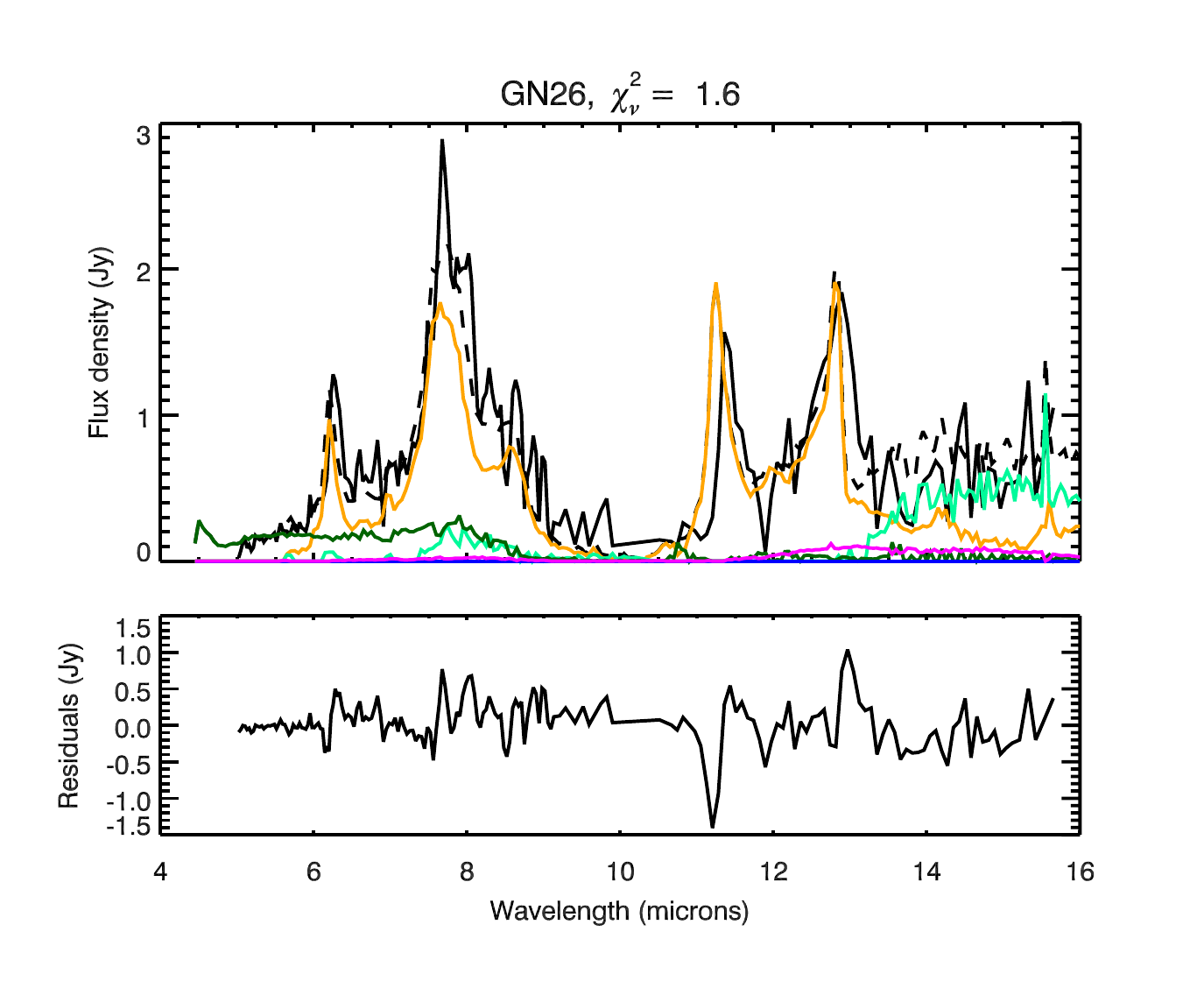} \\
\includegraphics[width=8.5cm]{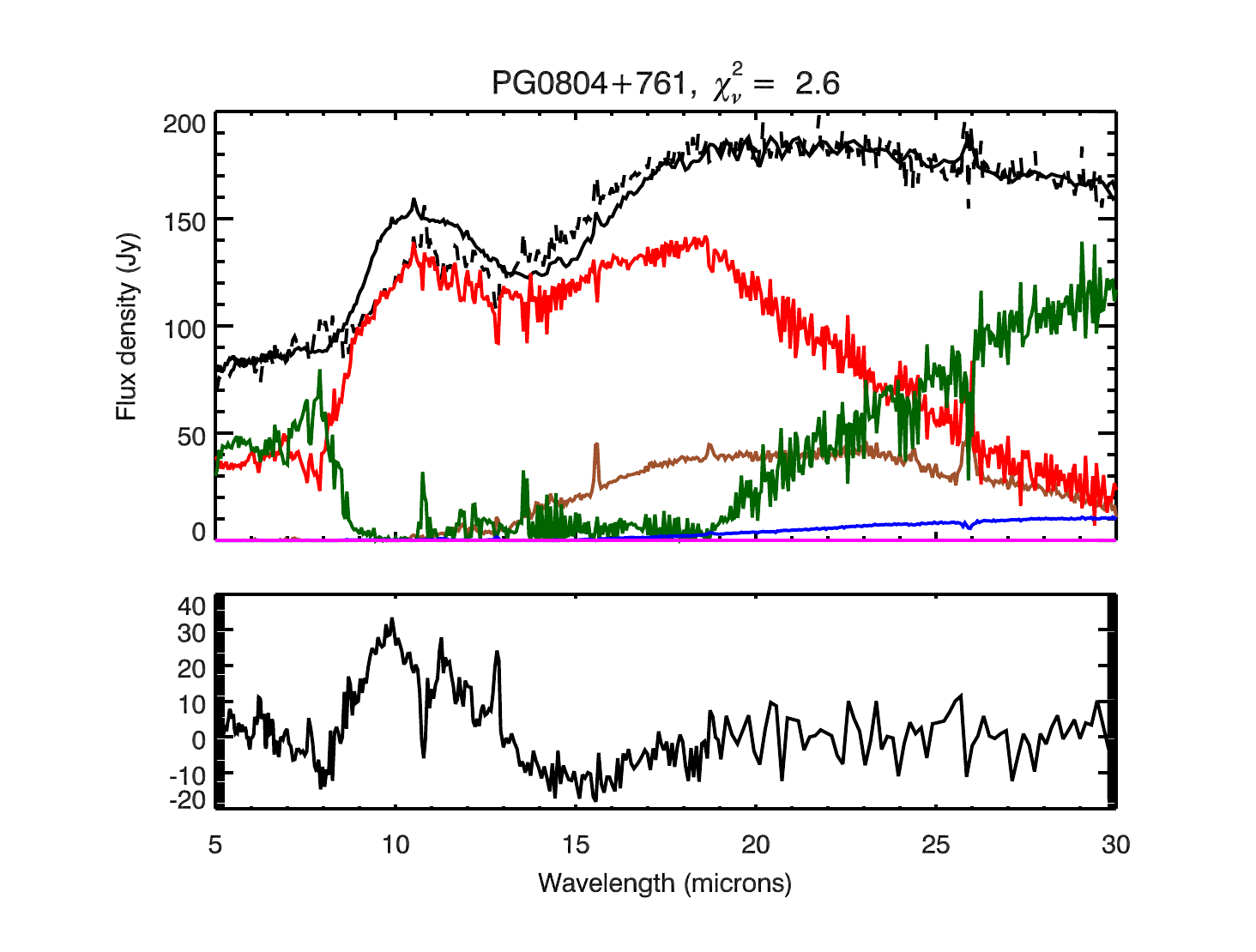} & \includegraphics[width=8.5cm]{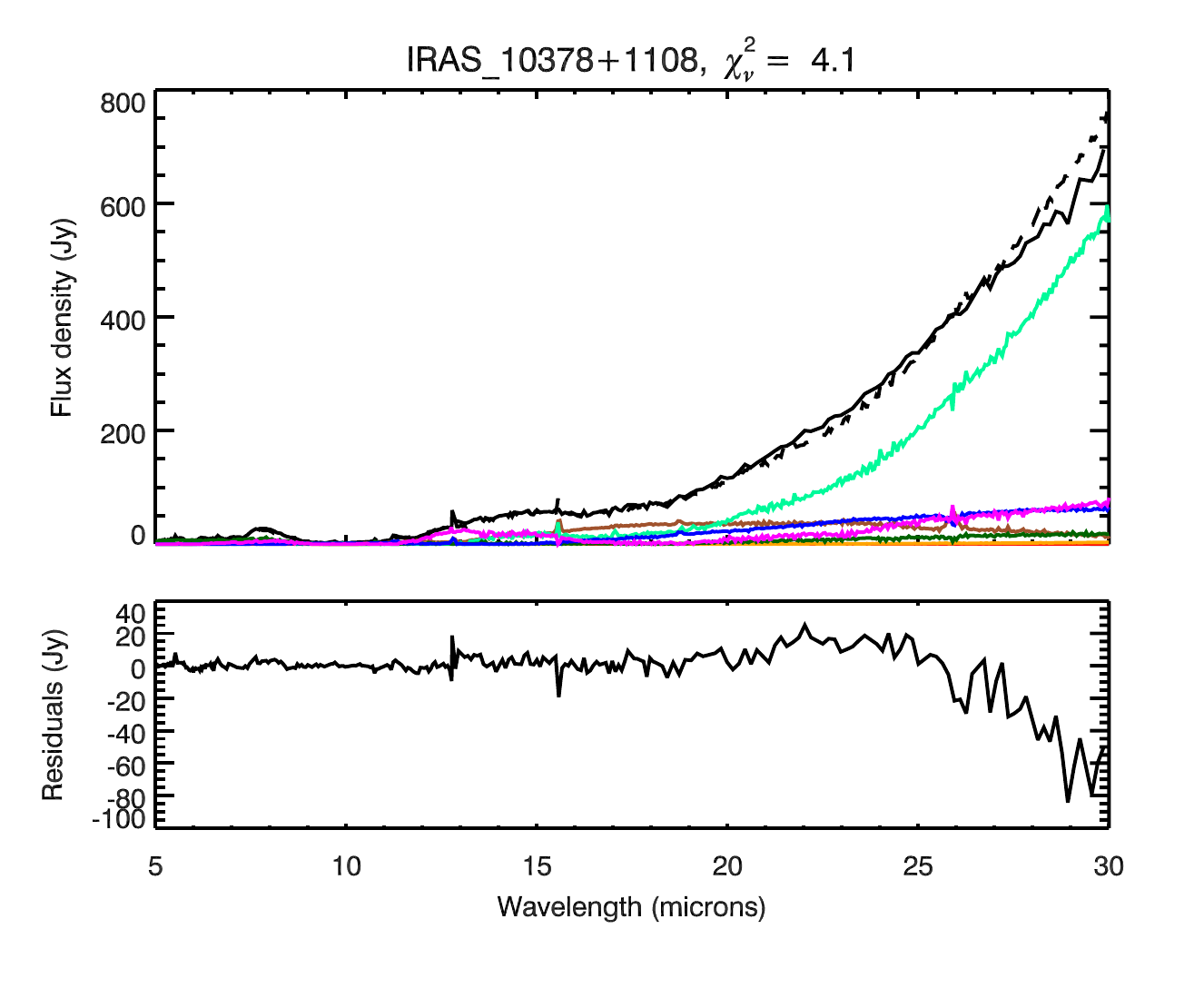} \\
\end{tabular}
\caption{Four additional example $NMF_{7}$ fits. LINER: 3C270, submillimeter galaxy: SMG GN26, quasar: PG0804+761, and ULIRG:IRAS 10378+1108 . Each spectrum is plotted as a black solid line and the NMF fit as black dashed line. The contribution from each component is also shown, with the same colour coding as in Figure \ref{7_components}}\label{example_gal_fits2}
\end{figure*}

\begin{figure*}
\begin{tabular}{c c}
\includegraphics[width=8.5cm]{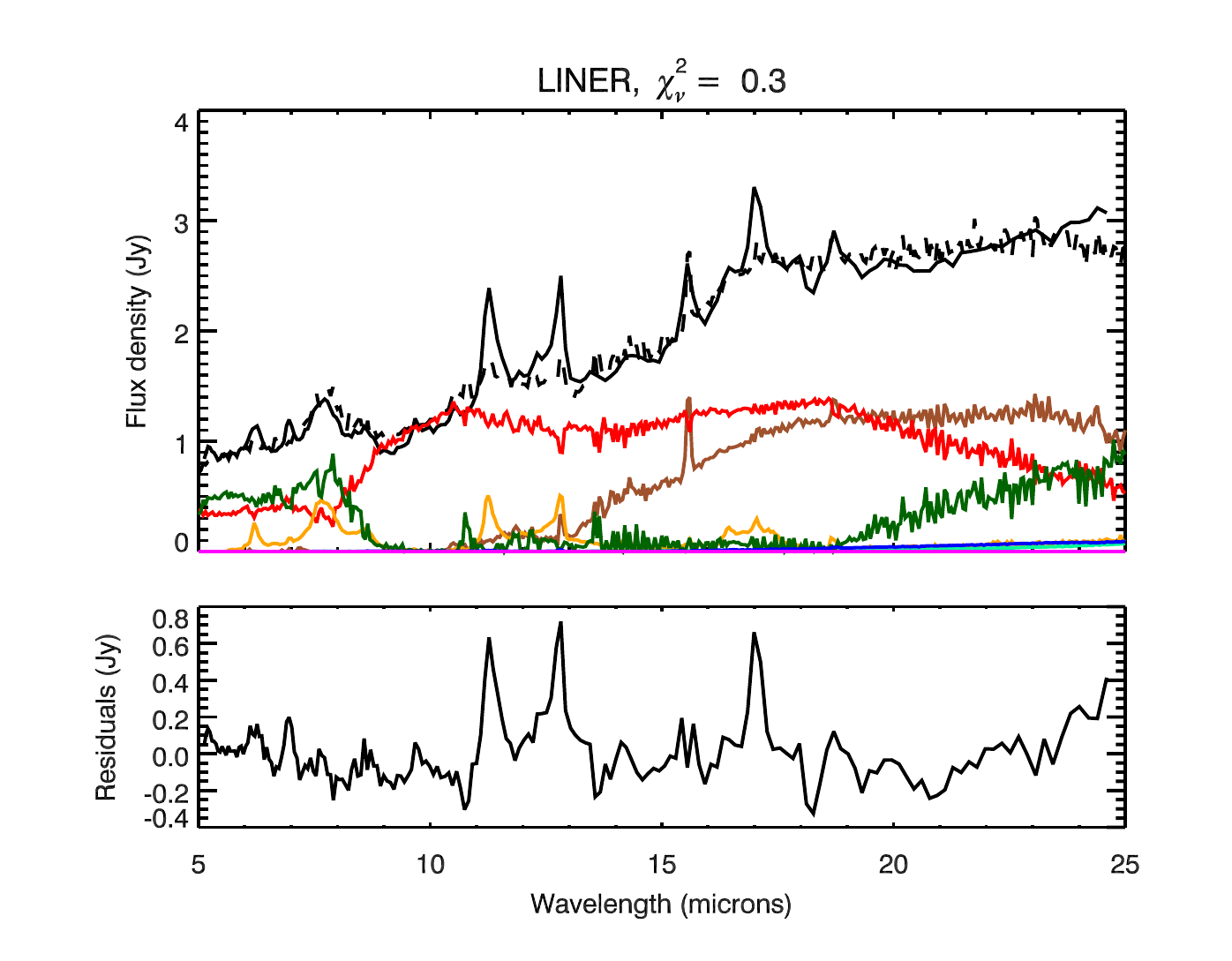} & \includegraphics[width=8.5cm]{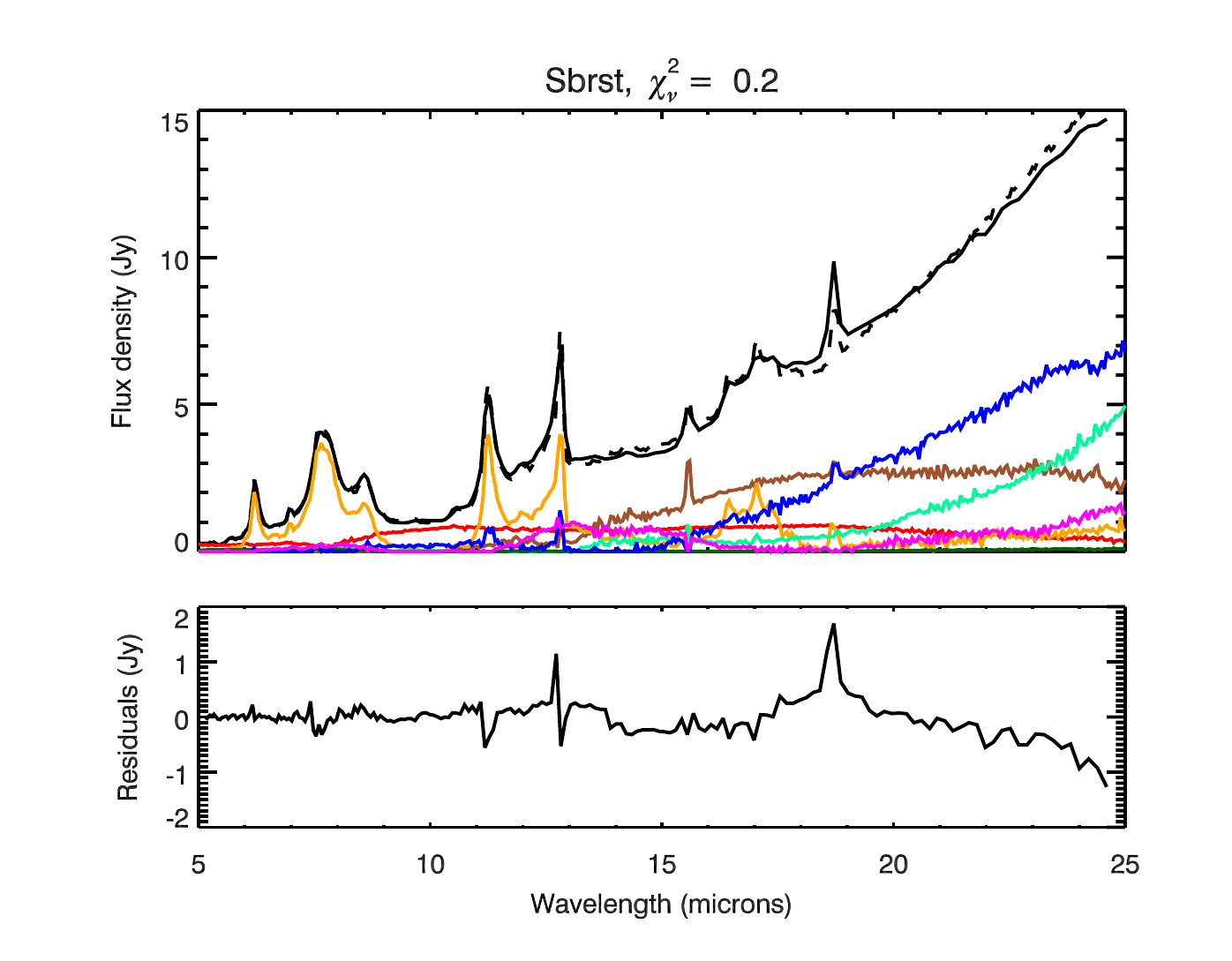} \\
\includegraphics[width=8.5cm]{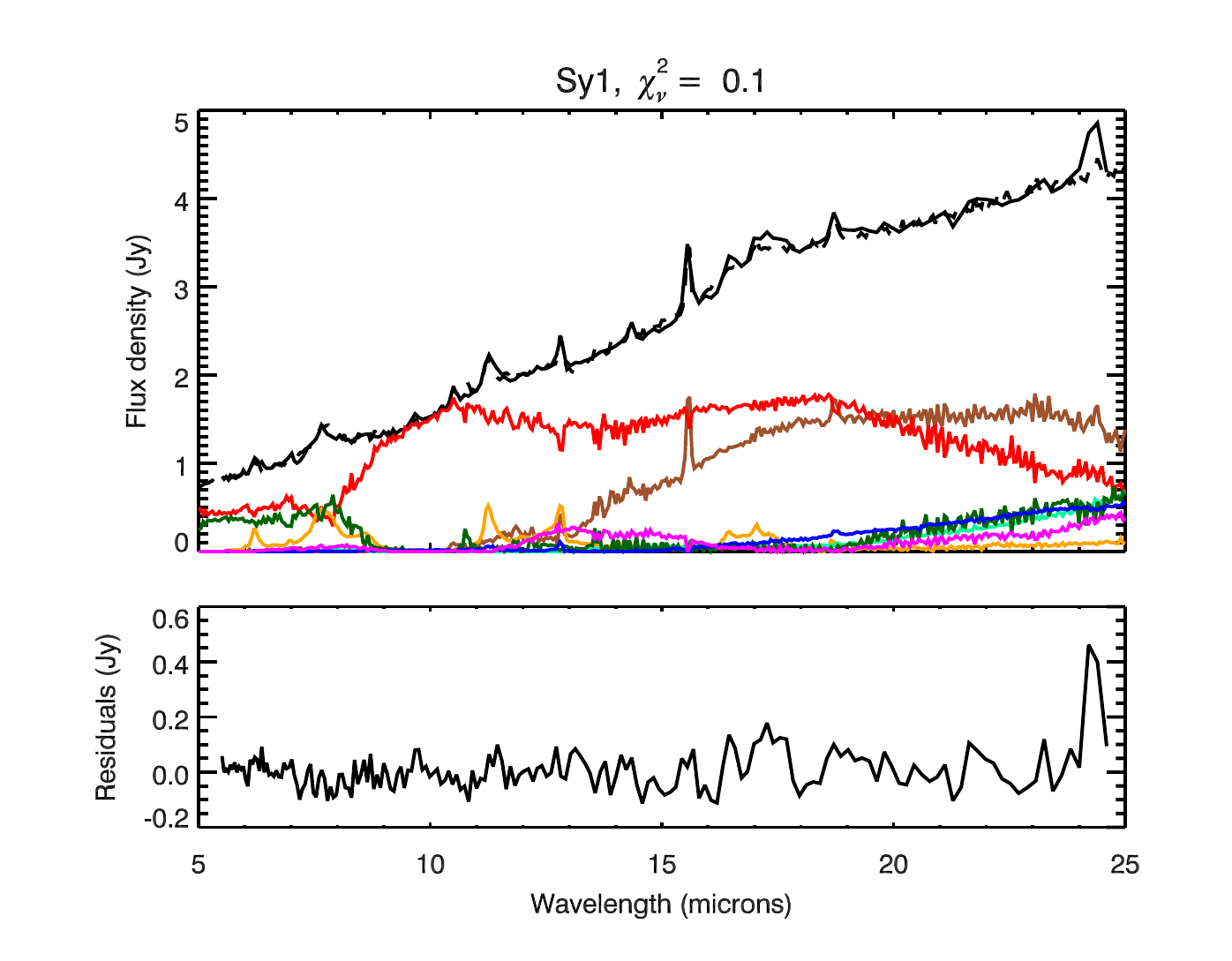} & \includegraphics[width=8.5cm]{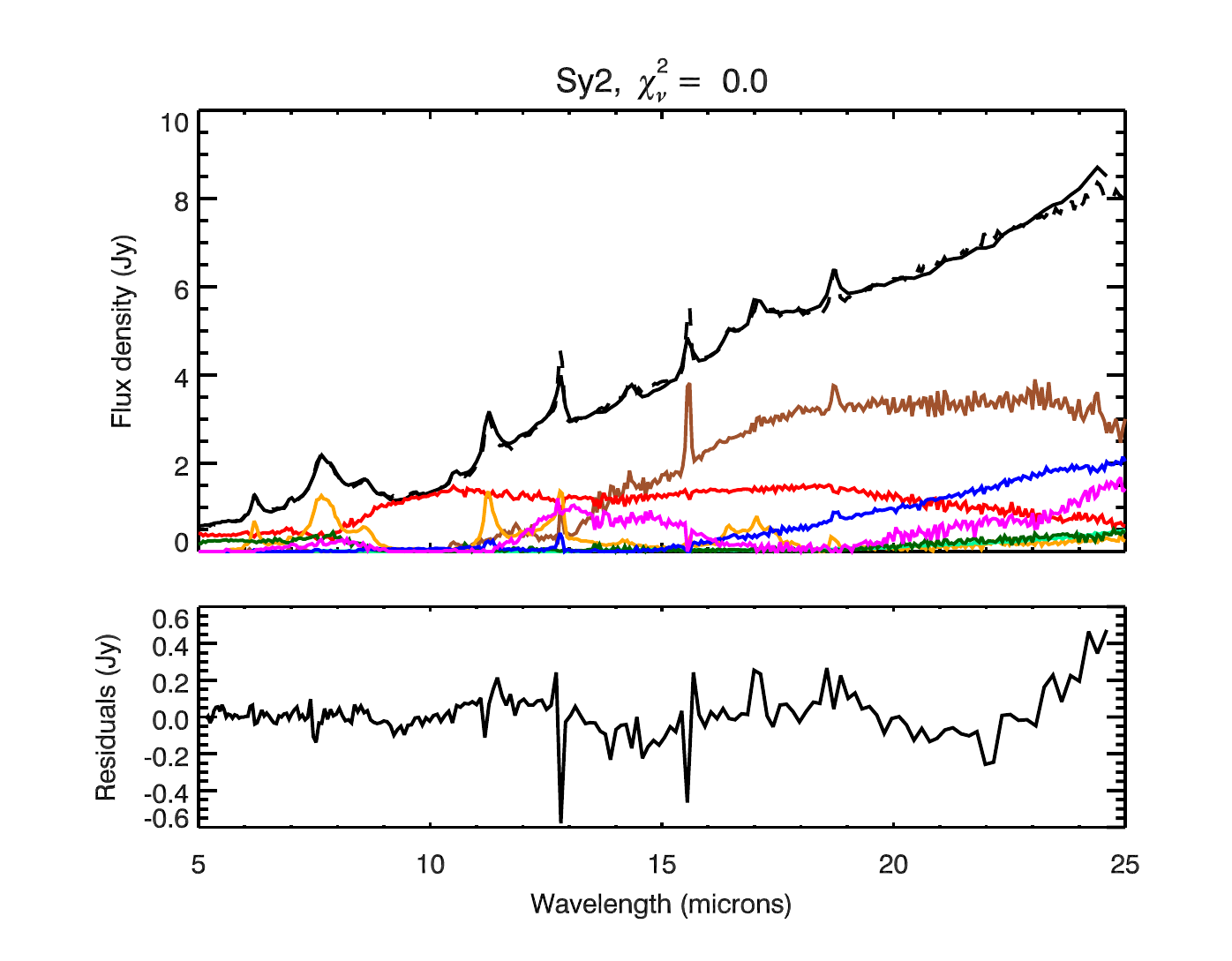} \\
\includegraphics[width=8.5cm]{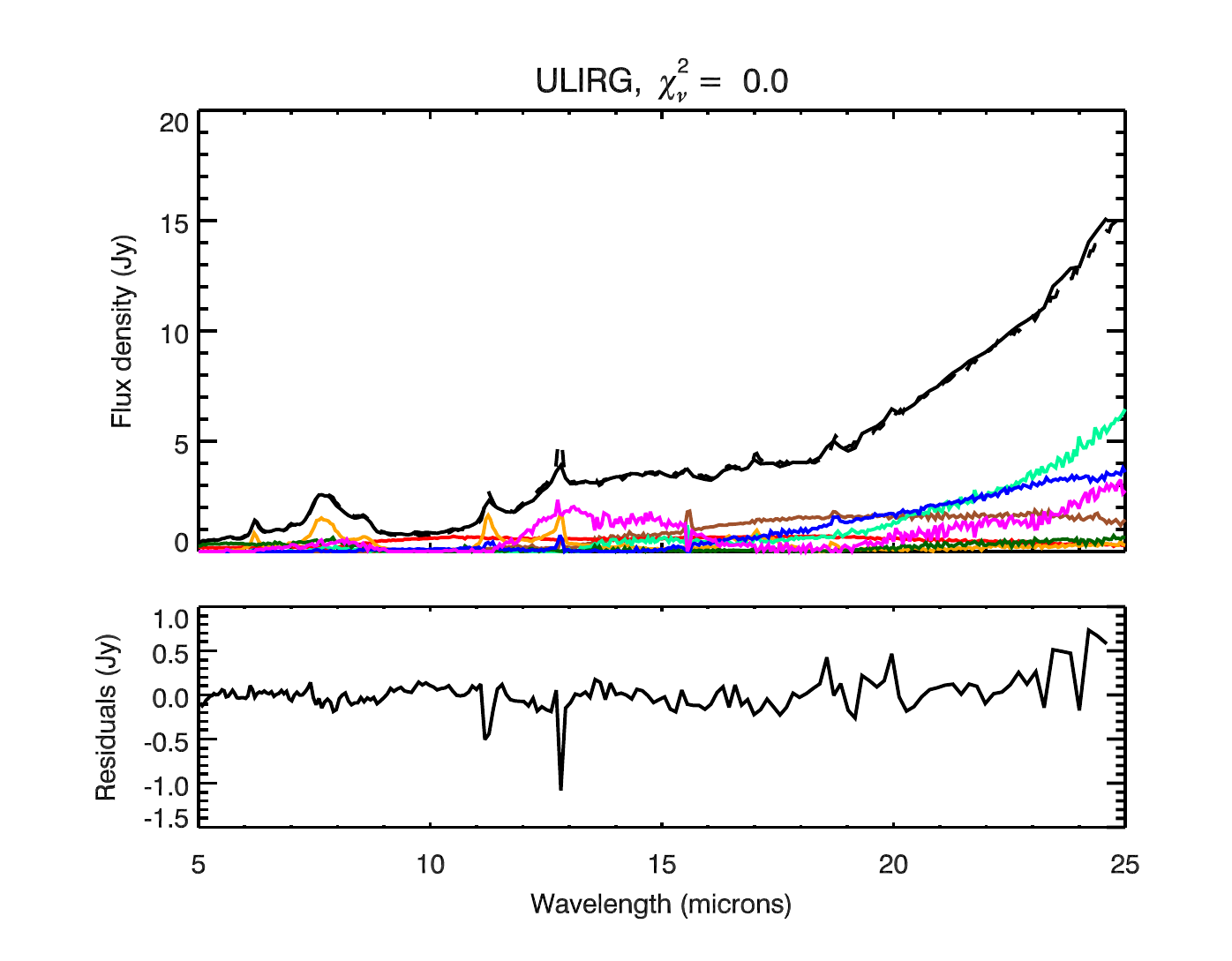} & \includegraphics[width=8.5cm]{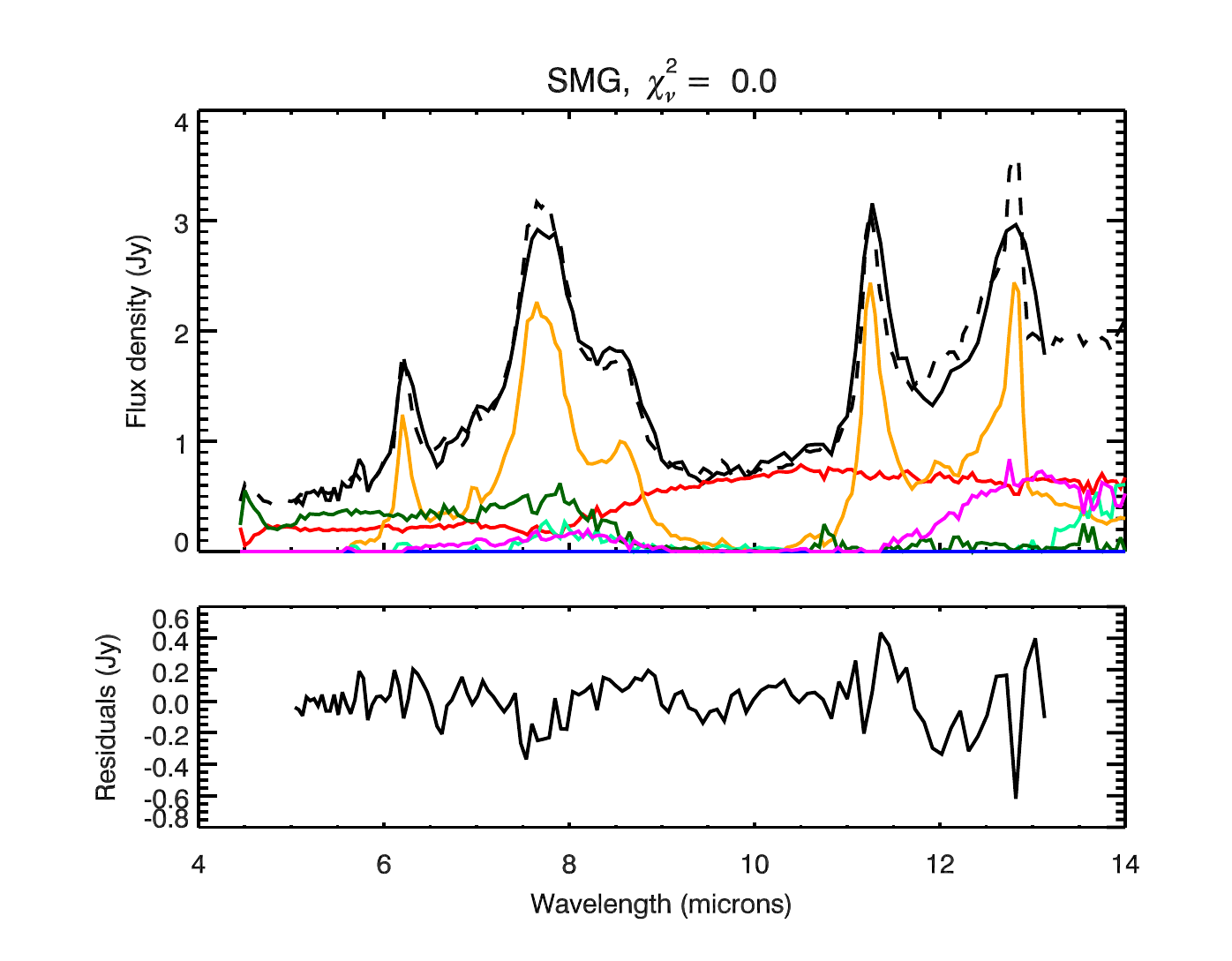} \\
\end{tabular}
\caption{$NMF_{7}$ fits to the Average templates from \protect\cite{Hernan-Caballero:2011} (Information on sample can be found in Table \ref{tabla_templates}). Each spectrum is plotted as a black solid line and the NMF fit as black dashed line. The contribution from each component is also shown, with the same colour coding as in Figure \ref{7_components}.}\label{example_av_fits}
\end{figure*}

\input{ATLAS_temp_table}

%
%
%
%
%
%%%%%%%%%%%%%%%%%%%%%%%%%%%%%%%%%%%%%
%
%% \bsp % ``This paper has been produced using the ...''
%
%\label{lastpage}

\end{document}

%% file: Introduction_v2.tex
\section{Introduction}

Spectra of the integrated mid-infrared (MIR) emission from galaxies contain a wealth of diagnostics that probe the origin of their MIR luminosity. For example, the main polycyclic aromatic hydrocarbons (PAHs) emission features found at 6.2, 7.7, 8.6, 11.3 and 12.7 $\mathrm{\mu m}$ are strong in objects where star formation activity contributes significantly to the mid-IR luminosity \citep{Genzel:1998, Laurent:2000}. The PAH features are either weak or absent for objects dominated by an active galactic nucleus (AGN) while emission lines with a high ionisation potential, for example the Neon fine structure line [Ne V] 14.3 $\mathrm{\mu m}$, tend to be strong in the presence of an AGN \citep{Genzel:1998, Sturm:2000}. Ratios of other fine structure lines such as [Ne III] $\mathrm{\mu m}$ 15.56 / [Ne II] 12.81 $\mathrm{\mu m}$ versus [S III] 33.48 $\mathrm{\mu m}$/[Si II] 34.82 $\mathrm{\mu m}$ have been shown to diagnose power source \citep{Dale:2006} as has the shape of the underlying mid-infrared dust continuum. \citep{Brandl:2006}.

Observations from the Infrared Space Observatory (ISO: \citep{Kessler:1996nl}), and the Infrared Spectrograph (IRS; \citep{Houck:2004pi}) on the Spitzer Space Telescope \citep{Werner:2004gm} allowed the MIR spectral features to be used as diagnostics of star formation and AGN activity. Combinations of the PAH emission lines, high to low excitation mid-infrared emission lines, silicate features and continuum measurements have been used as diagnostics for characterising the power source behind Ultraluminous Infrared Galaxies (ULIRGs) \citep{Genzel:1998, Rigopoulou:1999,Spoon:2007uq, Farrah:2007, Farrah:2008uk,Farrah:2009,Petric:2010}. 
 
However, diagnostics based on specific emission and absorption lines only focus on small parts of the spectrum, disregarding the information contained in the rest of the mid-infrared region. They can also be ambiguous. Dust and gas require ionising radiation to emit in the mid-IR, the source of the radiation is not important. For example, hot OB stars or an accretion disk around a supermassive black hole can both produce the [OIV] 25.9 $\mathrm{\mu m}$ emission line, as well as shocks \citep[e.g.][]{Lutz:1998}. The line ratios of fine structure lines can also be affected by the geometry of the emitting region and the age of a starburst, while the metallicity can affect PAH emission strength\citep[e.g.][]{Thornley:2000, Engelbracht:2005, Madden:2006, Wu:2006, Farrah:2007}. As a result, different diagnostics can give conflicting estimates for the contribution from star formation and/or AGN \citep[e.g.][]{Armus:2007, Veilleux:2009}.

Separation of spectral features from continuum and the mixing of neighbouring spectral features can also be problematic. For example, measurement of the 9.7 $\mathrm{\mu m}$ silicate feature requires different methods depending on the strength of the 8.6 and 11.2 $\mathrm{\mu m}$ PAH emission lines \citep{Spoon:2007uq}.  

An alternative method for identifying the power source is to decompose the spectra with AGN and starburst spectral templates. These templates tend to be a spectrum from a specific object (e.g. M82) or a mean spectrum of a number of similar object types. \cite{Pope:2008} use a combination of the M82 spectrum, average spectral template of starburst galaxies \citep{Brandl:2006} and a power law to decompose the IRS spectra of 13 high redshift submillimeter galaxies. \cite{Valiante:2009} fit IRS spectra across the range 5.5-6.85 $\mathrm{\mu m}$ with a combination of the M82 spectrum and a linear approximation for the AGN continuum. \cite{Alonso-Herrero:2011} use the \citep{Brandl:2006} starburst template and CLUMPY radiative transfer models for AGN to decompose the IRS spectra of 53 LIRGs into starburst and AGN components. Using average starburst templates is both simplistic and problematic. Prior theoretical prejudices drive the choice for what objects are used for the average templates, and they may be contaminated by AGN emission. The same is true for AGN average spectral templates.

With the public release of all low resolution Spitzer/IRS spectra by the Cornell Atlas of Spitzer/IRS sources (CASSIS)\citep{Lebouteiller:2011}\footnote{The Cornell Atlas of Spitzer/IRS Sources (CASSIS) is a product of the Infrared Science Center at Cornell University, supported by NASA and JPL.}, we are now in a better position to investigate the role played by star formation and AGN with more sophisticated techniques. In this paper we use a multivariate analysis technique to blindly learn the fundamental MIR spectral components, which we interpret as different physical environments within galaxies. Learning the MIR spectral shape of physical environments, allows the whole MIR wavelength range to be used as a diagnostic. The spectral components also provide an alternative to average spectral templates. 

A subclass of multivariate analysis techniques include matrix factorisation algorithms. The techniques are often associated with pattern recognition and blind source separation \cite{Lee:2001}. Algebraically, the algorithms approximate a data matrix by two simpler matrices: a weight matrix and component matrix. Common factorisation techniques include Singular Value Decomposition, Principal Component Analysis and Independent Component Analysis. The different techniques use different assumptions to carry out the factorisation, resulting in different weights and components. As multivariate datasets of spectra have become more prevalent, techniques such as Principal Component Analysis (PCA) have been applied to astronomical problems.  PCA has already been used for spectral classification of optical galaxies \citep[e.g.][]{Connolly:1995,Bromley:1998,Taghizadeh-Popp:2012}. PCA has also been successfully applied to the IRS spectra of local ULIRGs \citep{LingyuPCA,Hurley:2012}. 

The weights and spectral templates derived with PCA can be both positive and negative. Spectral reconstruction involves both addition and cancellation of spectral features. As a result, the PCA templates are inherently difficult to interpret physically.

A relatively new matrix factorisation technique, Non-negative matrix factorisation (NMF;\cite{Lee:1999}) can be thought of as PCA but with non-negative constraints on weights and templates. The constraints make reconstruction a purely additive process which more closely resembles emission in the mid-infrared. The first application of NMF to astronomy was carried out by \cite{Blanton:2007} who adopted the \cite{Lee:2001} NMF algorithms and applied it to optical spectra and photometry. It has also been used as a blind source separation algorithm on the IRS spectra of galactic photo-dissociation regions \citep{Berne:2007, Rosenberg:2011}.

This paper presents the first NMF analysis on mid-infrared galaxy spectra. We use spectra from the recently released Cornell Atlas of Spitzer/IRS sources (CASSIS) \citep{Lebouteiller:2011}. Our paper provides the first large scale statistical analysis of the IRS spectra to date using the NMF algorithm. Section \ref{sec:cassis} describes the CASSIS database and data reduction. In Section \ref{sec:matrix}, we describe the suitability of matrix factorisation to IRS spectra, and give details on the NMF algorithm. In section \ref{sec:results} we present our results and in Section \ref{sec:conclusions} our conclusions.  We assume a spatially flat cosmology with $H_{0} = 70 \mathrm{km s^{-1} Mpc^{-1}}, \Omega = 1$, and $\Omega_{m} = 0.3.$

%% file: CASSIS.tex
\section{The Data}\label{sec:cassis}
\subsection{CASSIS}
We use spectra from the Cornell Atlas of Spitzer/IRS sources (CASSIS) \citep{Lebouteiller:2011}. The atlas contains sources observed in low resolution mode with the Infrared Spectrograph (IRS;\cite{Houck:2004pi}) on board the \emph{Spitzer Space Telescope} \citep{Werner:2004gm}. IRS low resolution mode observations were made using two low-resolution modules, ShortÐLow and LongÐLow (hereafter SL and LL), covering 5.2-14.5 and 14.0-38.0 $\mathrm{\mu m}$ respectively. The modules also had a resolving power of $R \approx 60 - 120$ ($\approx 75\%$ of the observations) and an aperture size of $3.7 \times 57''$ for SL and $10.7 \times 168''$ for LL. The observations in the CASSIS database are first processed with the Basic Calibrated Data (BCD) pipeline from the \emph{Spitzer} Science pipeline (release S18.7.0.) and produces BCD frames. This removes electronic and optical artefacts. The BCD images are then processed using the CASSIS pipeline which carries out image cleaning, background subtraction, and spectral extraction. The pipeline algorithm is both automatic and flexible enough to handle different observations, from barely detected sources to bright sources and from point-like to somewhat extended sources.

\subsection{Sample}
\begin{figure}
\includegraphics[width=8.0cm]{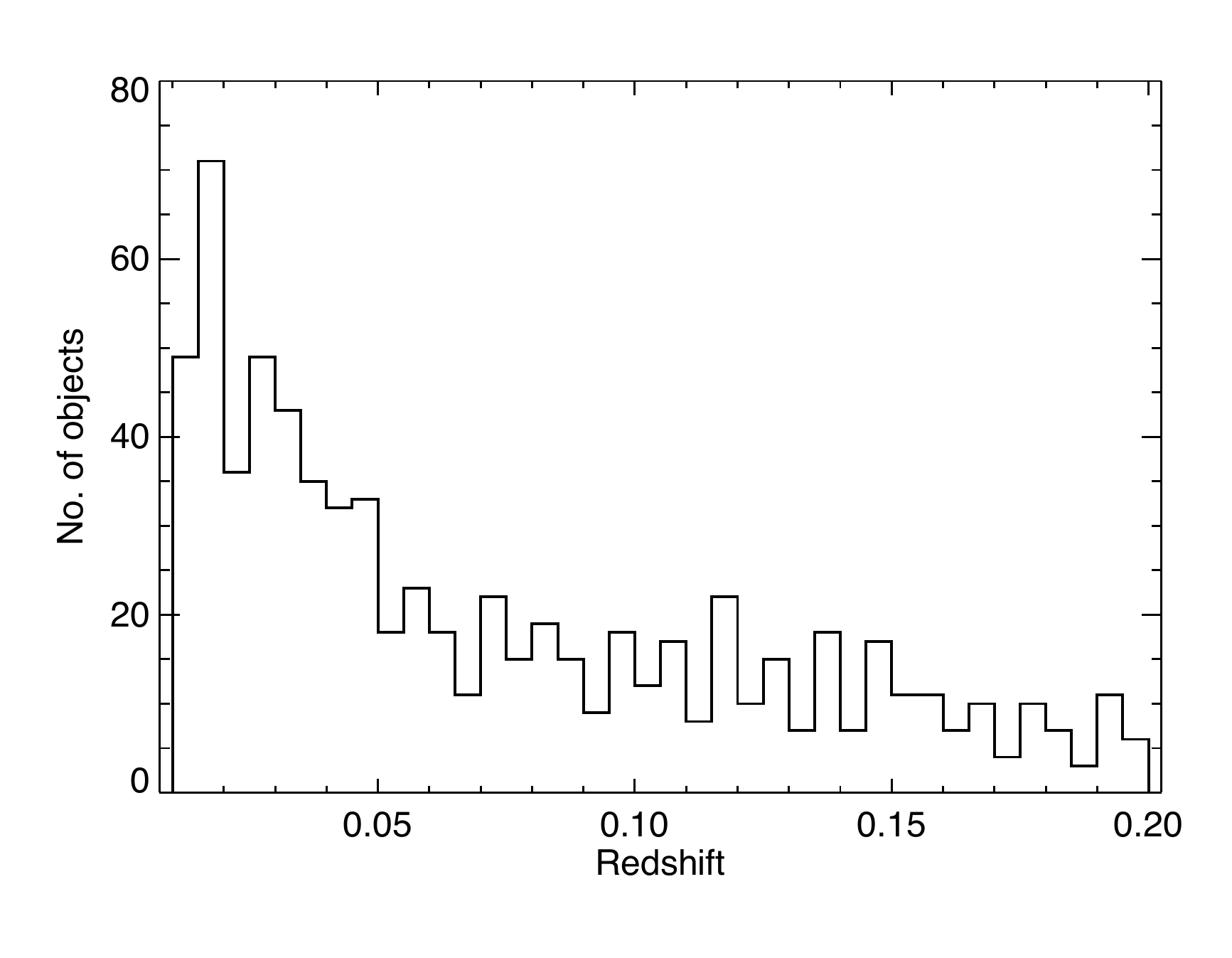}
\caption{The redshift distribution for the sample selection we apply the NMF algorithm to.}\label{fig:redshift_dist}
\end{figure}

The current version of CASSIS (version 4) contains 11304 distinct sources. 2118 of those distinct sources have known spectroscopic redshifts taken from NASA/IPAC Extragalactic Database (NED\footnote{\url{http://nedwww.ipac.caltech.edu/}}). We make the additional redshift cut $(0.01<z<0.2)$. The lower limit prevents contamination from Galactic and local group sources while the upper limit ensures we sample approximately the same wavelength range for each object. The redshift cut gives us a sample size of 893. We note that the redshifts within CASSIS, have been collected heterogeneously, biasing our sample by the parent redshift surveys. Because objects in our sample are at low redshift and span many programs, they likely span most or all IR luminous object types in the local Universe. Therefore, while a small degree of bias is inevitable, we do not consider that it is significant enough to significantly affect our results. We also only use objects with both SL and LL data. This reduces our sample size down to 729 objects. The redshift distribution for the 729 objects can be seen in Figure \ref{fig:redshift_dist}.

\subsection{Stitching}
Observations using data from both SL and LL spectral modules can suffer from mismatching due to telescope pointing inaccuracy or if a source is extended in SL and not in LL. The mismatching causes the spectra from one of the modules (normally the SL) to have lower flux calibration than the other. Correcting the mismatch is inherently difficult as the data from the overlap between the two modules can suffer from the '14 micron teardrop' (see IRS instrument handbook, \footnote{\url{http://irsa.ipac.caltech.edu/data/SPITZER/docs/irs/}}) , leaving a small gap at around 13-14 $\mathrm{\mu m}$. 

We correct for the mismatch using a simplified version of our NMF technique. For the first step, we generated two sets of templates, one using SL data and the other using LL data. The distribution in redshift causes the mismatch region to occur at different rest frame wavelengths for different objects. This ensures at least one template set covered the mismatch region for each object. We then fitted the template set to a region of width 7 $\mathrm{\mu m}$,  centred on the mismatch area. Wavelength points associated with PAH and Neon emission lines were removed to prevent strong line strengths from distorting the fits. We carry out the fit for different scalings applied to the SL data. The scaling factor value that gives the lowest $\chi^{2}$ is chosen as the scaling correction. Having stitched the spectra using both SL and LL template sets, we then generated our initial NMF sets for the entire spectral range. We then re-stitch the spectra with the new NMF set. Additional spectra used for analysis in this paper are also stitched with our final $NMF_{7}$ set, introduced in section \ref{sec:results}.

\subsection{Normalisation}
The NMF analysis requires all spectra to be normalised to a standard value to prevent sources with higher flux, biasing the algorithm. We normalise all the spectra by the average flux across the restframe wavelength range of $7-20 \mathrm{\mu m}$. We choose this range as it is common to all sources with both SL and LL data. 

%% file: Matrix_factorisation.tex
\section{Matrix Factorisation}\label{sec:matrix}
Analysis of spectra from Spitzer's IRS has tended to be done using diagnostics based on only a few of the specific features \citep[e.g.][]{Sajina:2007, Pope:2008, Alonso-Herrero:2012}. For example, \cite{Spoon:2007uq} introduced a classification scheme based on the 6.2$\mathrm{\mu m}$ PAH line and $10\mathrm{\mu m} $ silicate feature. Quantifying the contribution from star formation and AGN has also been carried out using fine structure lines, for example the [OIV]/[NeII] and [NeV]/[NeII] line ratios versus the 6.2$\mathrm{\mu m}$ PAH equivalent width. \citep[e.g.][]{Armus:2007,Petric:2010}

In essence, line diagnostic analyses are carrying out a crude compression by using only small parts of the spectrum to describe each object (e.g. the 6.2$\mathrm{\mu m}$ feature). Matrix factorisation techniques provide an alternative approach to compression by transforming data from wavelength space to one that better captures the variance in the dataset. As a result, classification or quantification of properties such as star formation is carried out considering a greater wavelength range. 

Algebraically, matrix factorisations find a linear approximation to a data matrix X such that $\mathrm{X} \thickapprox \mathrm{WH}$, or:
\begin{equation}
\mathrm{X}_{i\mu} \thickapprox (\mathrm{WH})_{i\mu} = \sum_{a=1}^{r} \mathrm{W}_{ia}\mathrm{H}_{a\mu}
\end{equation}

Where, $i$ is object index, $\mu$ is wavelength and $a$ is component index. The matrix H can be thought of as a set of $r$ components that represent latent structure explicit in the dataset, and W are a set of weighting coefficients. Each object in the dataset can now be approximated by a linear combination of the derived components, H.

Different matrix factorisation techniques use different assumptions to carry out the approximation. Independent component analysis (ICA) assumes the derived components (H) are independent. 
Principal component analysis (PCA) models the dataset as a multivariate Gaussian distribution in wavelength space and finds the orthogonal components of the Gaussian. Non-Negative Matrix Factorisation (NMF) assumes the data, weights and components are all non-negative, but makes no assumption on the distribution of the data or correlation between derived components.

\subsection{Matrix factorisation of spectra}
By applying linear matrix factorisation techniques to the mid-infrared spectra of galaxies, we are assuming mid-infrared spectra of galaxies, $F(\lambda)$, can be modelled as a linear combination of components. Ideally the components would relate to physical regions, for example a star forming region ($T_{SF}$), an active galactic nuclei torus ($T_{AGN}$), a molecular cloud ($T_{MC}$) or diffuse dust component ($T_{C}$). A spectrum for a galaxy would then simply be:
\begin{equation}
F(\lambda) = a\cdot T_{SF}(\lambda) + b\cdot T_{AGN}(\lambda) + c\cdot T_{MC}(\lambda) + d\cdot T_{C}(\lambda)
\label{eq:linear}
\end{equation}
Where, $a, b, c$ and $d$ are the relative weights for each component.

For the above model, ICA is not suitable as the components are unlikely to be independent, for example AGN and star formation are believed to be triggered by similar mechanisms such as mergers \citep[e.g.][]{Sanders:1988}, and are likely to be connected through feedback processes \citep[e.g.][]{Farrah:2012,Rovilos:2012}.
 
PCA has already been applied to the mid-infrared spectra of ULIRGs \citep[e.g.][]{LingyuPCA, Hurley:2012}. Algebraically, PCA calculates the eigenvectors of the covariance matrix. For spectra, the principal components represent the principal variations from a mean spectral template. The components are therefore allowed to have features which are positive and negative, and are also allowed to have a negative weighting when fitting objects. The freedom to be both positive and negative does not mimic the process of emission in the MIR, resulting in components that are inherently difficult to interpret. By their nature, the principal components have a statistical rather than physical interpretation. Therefore, although PCA can successfully reduce dimensionality of spectra for classification from known objects, it is not suitable for our model.

The non-negative constraint of NMF more closely reflects the physical process of emission in the mid-infrared, which does not suffer from the same problems of absorption as other spectral ranges. As a result the NMF generated templates are more physically intuitive. 

NMF is therefore the most applicable matrix factorisation routine for our linear interpretation of galaxy emission. However, the situation is complicated by dust extinction. This introduces a non-linearity to the problem since extinction is multiplicative and exponential. 

\begin{equation}\begin{split}
F(\lambda) = (a\cdot T_{SF}(\lambda) + b\cdot T_{AGN}(\lambda) + c\cdot T_{MC}(\lambda) \\
+ d\cdot T_{C}(\lambda))e^{-f\cdot\tau(\lambda)}
\end{split}\label{eq:simple_ext}
\end{equation}
Where $f$ is the weight associated with extinction and $\tau(\lambda)$ can either be known or unknown. 

We can take the model one step further by allowing extinction to vary across all four components:
\begin{equation}\begin{split}
F(\lambda) = a\cdot T_{SF}(\lambda)e^{-f\cdot\tau(\lambda)} + b\cdot T_{AGN}(\lambda)e^{-g\cdot\tau(\lambda)} \\
+ c\cdot T_{MC}(\lambda)e^{-h\cdot\tau(\lambda)} + d\cdot T_{C}(\lambda)e^{-i\cdot\tau(\lambda)}
\end{split}\label{eq:complicated_ext}
\end{equation}
The weights for the extinction are $f, g, h$ and $i$.

We have explored the suitability of non-linear kernel based matrix factorisation algorithms \citep[e.g.][]{Zafeiriou:2010,Pan:2011} and found they are not suited for the non-linear behaviour described in equations \ref{eq:simple_ext} and \ref{eq:complicated_ext}. We discuss why in Appendix A. Current algorithms therefore restrict us to describe mid-infrared galaxy spectra as a set of linear components (e.g. equation \ref{eq:linear}) and NMF is the most appropriate matrix factorisation technique. 

The first application of NMF in astronomy was carried out by \cite{Blanton:2007} who updated the popular NMF multiplicative algorithm from \cite{Lee:2001} to include uncertainties and for heterogeneous datasets (e.g. optical spectra and photometric observations of galaxies at different redshifts). They also restricted the space of possible spectra to those predicted from high resolution stellar population synthesis models. We use the NMF algorithm from \cite{Blanton:2007} to identify and learn the mid-infrared sources that are common to galaxies in the CASSIS database. Unlike \cite{Blanton:2007}, we do not use any models as a prior for shape of the components, we use the algorithm to blindly learn the shape of our components.

\subsection{NMF algorithm}
As with PCA, the goal of NMF is to minimise a cost function. The most widely used is the squared approximation error described in \cite{Lee:2001}:
\begin{equation}
\chi^{2} = \sum_{i\mu}\left(\mathrm{X}_{i\mu} - \sum_{a}\mathrm{W}_{ia}\mathrm{H}_{a\mu}\right)^{2}
\label{lee_cost}
\end{equation}

Minimising equation \ref{lee_cost} requires some sort of numerical technique to find local minima. \cite{Lee:2001} presented 'multiplicative update rules' for $\mathrm{H}$ and $\mathrm{W}$. Upon each iteration, the rules are used to update $\mathrm{H}$ and $\mathrm{W}$ by a multiplicative factor whilst minimising equation \ref{lee_cost}. The algorithm implemented in \cite{Blanton:2007} altered the original multiplicative update algorithm of \cite{Lee:2001} for nonuniform uncertainties ($\sigma$). The cost function then becomes the weighted squared approximation error:
\begin{equation}
\chi^{2} = \sum_{i\mu}\left(\frac{\mathrm{X}_{i\mu} - \sum_{a}\mathrm{W}_{ia}\mathrm{H}_{a\mu}}{\sigma_{i\mu}}\right)^{2}
\label{blanton_cost}
\end{equation}

\cite{Blanton:2007} showed the multiplicative update rules for $\mathrm{H}$ and $\mathrm{W}$ are as follows:
\begin{equation}
\mathrm{W}_{ia} \leftarrow \mathrm{W}_{ia} \left( \sum_{\mu}\frac{\mathrm{X}_{i\mu}\mathrm{H}_{a\mu}}{\sigma_{i\mu}^{2}}\right) \left(\sum_{m\mu}\frac{\mathrm{W}_{im}\mathrm{H}_{m\mu}\mathrm{H}_{a\mu}}{\sigma_{i\mu}^{2}}\right)^{-1}
\label{W_update}
\end{equation}

\begin{equation}
\mathrm{H}_{a\mu} \leftarrow \mathrm{H}_{a\mu}\left( \sum_{i}\frac{\mathrm{W}_{ia}\mathrm{X}_{i\mu}}{\sigma_{i\mu}^{2}}\right)\left(\sum_{mi}\frac{\mathrm{W}_{ia}\mathrm{W}_{im}\mathrm{H}_{m\mu}}{\sigma_{i\mu}^{2}}\right)^{-1}
\label{H_update}
\end{equation}

The update rules in equations \ref{W_update} and \ref{H_update} are guaranteed to reduce the error, however the cost function in equation \ref{blanton_cost} is not necessarily convex therefore the algorithm may get stuck in a local minimum. We run the algorithm five times with different initial starting positions to check the solution is consistent.

Convergence can be evaluated by looking at the decrease in cost function across iterations and checking the solution has reached a minimum. In practise, we find 3000 iterations are enough for $\mathrm{H}$ and $\mathrm{W}$ to converge.

The number of components generated by NMF is a user input. Unlike PCA where the shape of the original components remain unchanged as more are added, the NMF components will not remain the same. We investigate the number of components required to constrain the data by generating 11 different NMF sets, containing from 3 up to 14 components. We define the following notation, $NMF^{x}_{y}$  to describe the $x$th component from an NMF set containing $y$ components.

\subsection{Bayesian Evidence}
To determine the minimum number of components that are justified by the data, one should calculate the Bayesian evidence ($E$).
\begin{equation}
E \equiv \int L(\theta) \pi(\theta) d\theta
\end{equation}

The evidence can be thought of as the average likelihood, $L(\theta)$, over all of the prior, $\pi(\theta)$, parameter space, $d\theta$, of a given model and automatically implements Occam's razor, i.e. simpler models are preferred unless simplicity can be traded for greater explanatory power. 

There are two ways in which one could calculate the Bayesian evidence for our setup. The first would be to calculate the evidence for the NMF algorithm, where the number of parameters is equal to the number of elements in both $\mathrm{H}$ and $\mathrm{W}$. This approach would be the most appropriate if comparing the suitability of NMF to other matrix factorisation techniques, the integral however becomes highly multidimensional making the calculation numerically challenging. Alternatively, if NMF is the most appropriate algorithm to our problem, then we can assume that the components are correct. The number of parameters is then equal to the number of elements in $\mathrm{W}$, i.e. the number of components. 

We choose the later approach as we have already chosen NMF as the most appropriate algorithm to our problem and are not comparing alternative procedures.

We calculate the evidence by using the nested sampling routine, MULTINEST \citep{Feroz:2008} to re-fit the CASSIS sample with different NMF sets. MULTINEST is a Bayesian inference tool which calculates the evidence and produces posterior samples from distributions with (often an unknown number of) multiple modes and/or degeneracies between parameters. Nested Sampling \citep{Skilling:2004} is a Monte Carlo technique that randomly samples from the prior space, and zooms in on areas of higher likelihood during successive iterations.

We fit every galaxy with component sets $NMF_{3}$ to $NMF_{14}$ and their respective repeats. For every repeat, we calculate the median evidence of the sample. The main uncertainty on our evidence values comes from the difference in NMF sets across repeats (i.e. the convergence on slightly different local minima by the NMF algorithm). To quantify the uncertainty on our evidence values, we calculate the mean and standard deviation evidence values from the 5 repeats, as a function of number of components. 

%% file: Templatesv3.tex
\section{Results}\label{sec:results}

\begin{figure}
\includegraphics[width=8.5cm]{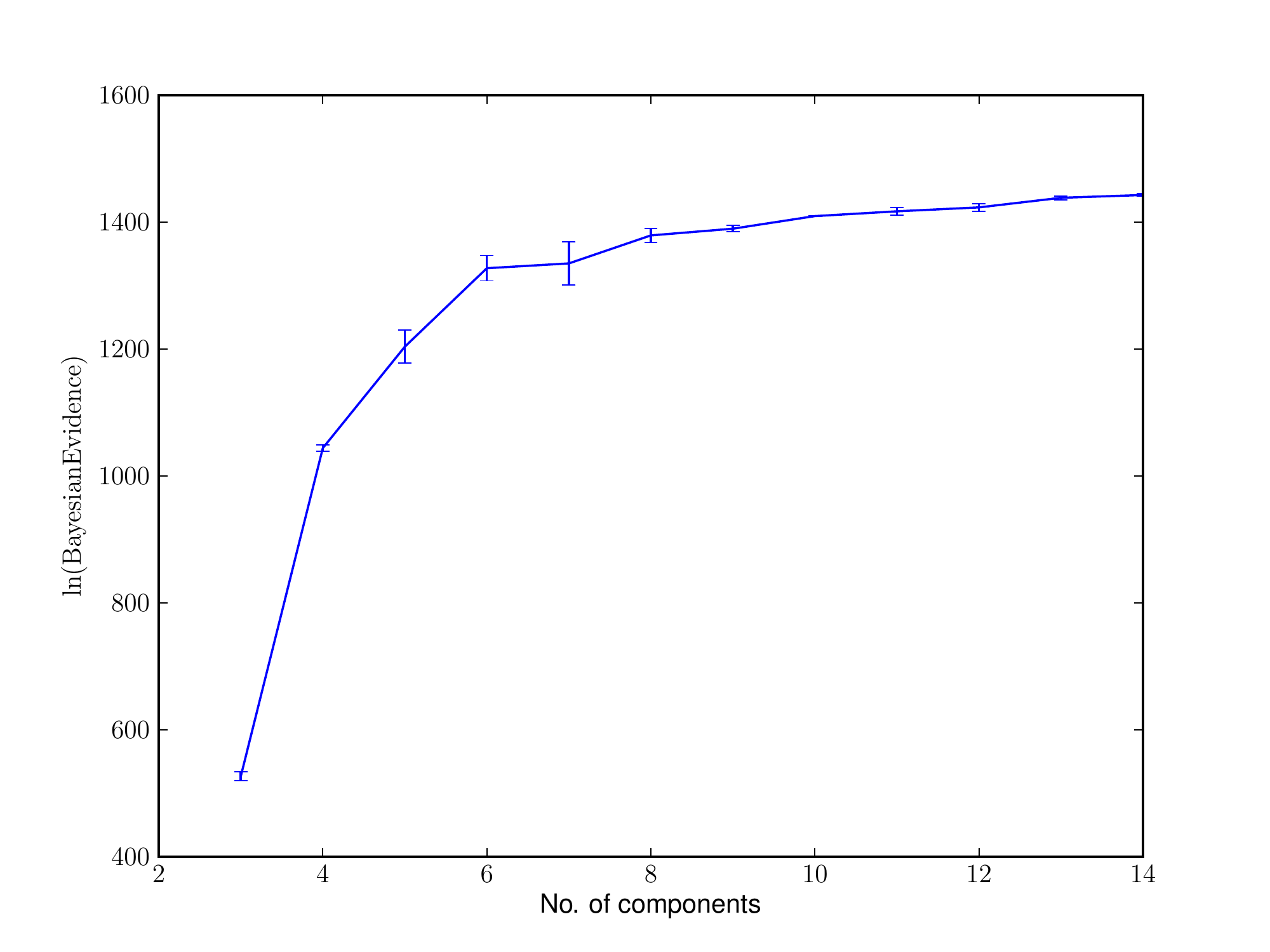}
\caption{The Bayesian evidence as a function of number of components. For each NMF set, we run the algorithm 5 times and calculate the median evidence value of the entire galaxy sample. We plot the mean and standard deviation of the 5 repeats.}\label{Bayes_ev}
\end{figure}

\subsection{Number of Components} 
As discussed in the previous section, we would like to quantify how many components are required by the data. Figure \ref{Bayes_ev} shows the mean and standard deviation for the Bayesian evidence values from 5 repeats, as a function of number of components. The Bayesian evidence should start decreasing as the number of components exceeds the optimum number needed to constrain the data. We see no turnover, indicating there is not an obvious, optimum NMF set below 14 components. We note however a slight levelling of at 7 components before increasing again beyond 8.

We have also looked at the ratio of evidence values between consecutive NMF sets. The ratio, referred to as the Bayes factor ($K$), is used as a measure for a Bayesian version of classical hypothesis testing. We use the Jeffreys scale to interpret $K$. A value of $K<1$ indicates the more complicated model is preferred, $K=1-3$ as barely worth mentioning, $K=3-10$ indicates substantial support for the simpler model, while $K=10-30$ is strong, $K=30-100$ is very strong, and $K > 100$ is considered decisive. Using the Jeffreys scale, we find more than 14 components are needed to reconstruct spectra within the uncertainties. However, we note that $K$ begins to level off after $6/7$, indicating that although more complicated component sets are preferred, the gain in increasing the number of components is beginning to decrease. 

Ideally, we would calculate the Bayesian evidence and Bayes factor beyond $NMF_{14}$. However, calculating evidence for highly multidimensional parameter spaces becomes computationally challenging. We have qualitatively examined NMF sets where number of components $>$ 14. As an example, in Figure \ref{NMF30} we show the NMF components for $NMF_{30}$. Interpreting a many-component NMF set such as $NMF_{30}$ becomes challenging as signatures begin to separate out into several components, whose physical interpretation is not clear. 

\begin{figure*}
\includegraphics[width=18cm]{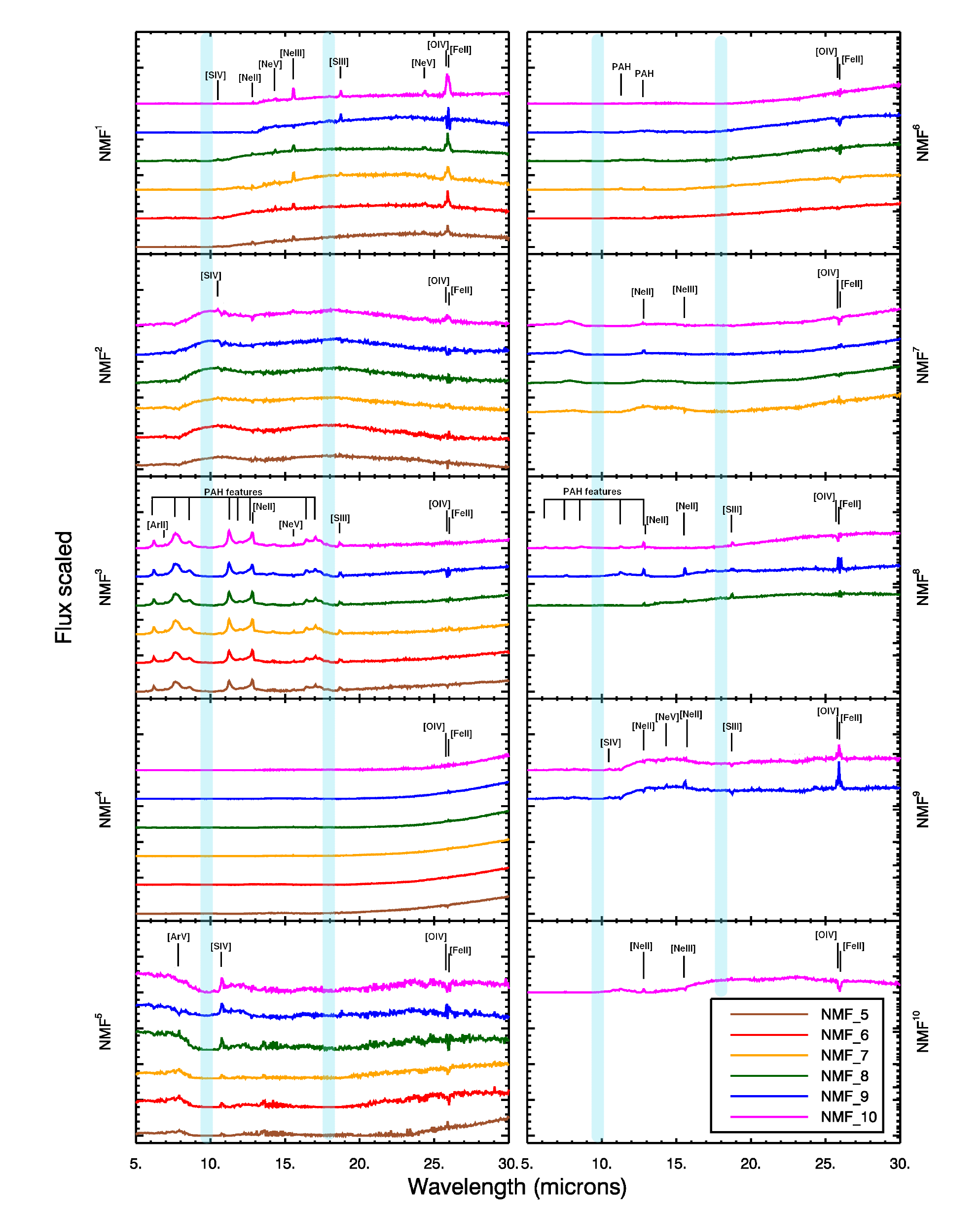}
\caption{The derived NMF spectral components for sets $NMF_{5}$-$NMF_{10}$. Each NMF set is colour coded, with components ordered by similarity. For example, the five components of $NMF_{5}$ are the five brown spectra. Prominent spectral features in each component are also labelled and regions affected by broad silicate absorption and emission are highlighted in light blue.}\label{fig:template_plots}
\end{figure*}

We also note that the Bayesian evidence calculation could be influenced by two fundamental factors. The first is the use of uncertainties associated with IRS spectra, which have often been underestimated below the observed variation between individual nod positions on the IRS, as described in Chapter 7 of the IRS Instrument Handbook \footnote{\url{http://irsa.ipac.caltech.edu/data/SPITZER/docs/irs/}}. As a result, our model selection may be too conservative. The other problem comes from the suitability of the NMF algorithm to the non-linear behaviour associated with extinction. We have carried out a simple simulation to show how extinction could be a factor in driving our linear methods to more templates than might be required by underlying physical conditions. Details can be found in Appendix B.

We have investigated how many components are needed in a quantitative manner. For the rest of this paper we investigate the how many components are needed qualitatively, by examining some of the simpler NMF component sets, limiting our investigation to $NMF_{5}$-$NMF_{10}$. 

\subsection{Analysis of $NMF_{5}$ to $NMF_{10}$}
Figure \ref{fig:template_plots} shows each spectral component for sets $NMF_{5} - NMF_{10}$. We have ordered the components so that similar components appear in the same order. We note the ordering of components given by NMF is unimportant.

The NMF sets in Figure \ref{fig:template_plots} show that many of the components remain similar, despite an increase in the allowed number of components. 

The first component contains a dust continuum which peaks at around 24 $\mathrm{\mu m}$ and contains emission from the Sulphur line [SIV] at 10.51 $\mathrm{\mu m}$, the 12.8, 15.6 and 24.3 $\mathrm{\mu m}$ Neon lines and Oxygen line [OIV] at 25.89 $\mathrm{\mu m}$, all of which are associated with a hot ionised gas source. The continuum in the component from $NMF_{9}$ and $NMF_{10}$ varies from the others in that continuum does not start until 13 $\mathrm{\mu m}$. This coincides with the appearance of the ninth and tenth components which show similar features. The hot dust continuum peaks at a wavelength similar to that of AGN tori, while the hot ionised gas emission lines have also typically been associated with AGN. The appearance of both in one component is consistent with the idea they are correlated.

The second component shows silicate emission features at 10 and 18 $\mathrm{\mu m}$ due to stretching and bending of the Si-O and O-Si-O bonds respectively. Silicate emission is typically associated with emission from very hot dust, found on the inner surface of AGN tori or narrow line regions \citep{Mason:2009}.

\begin{figure*}
\includegraphics[width=18cm]{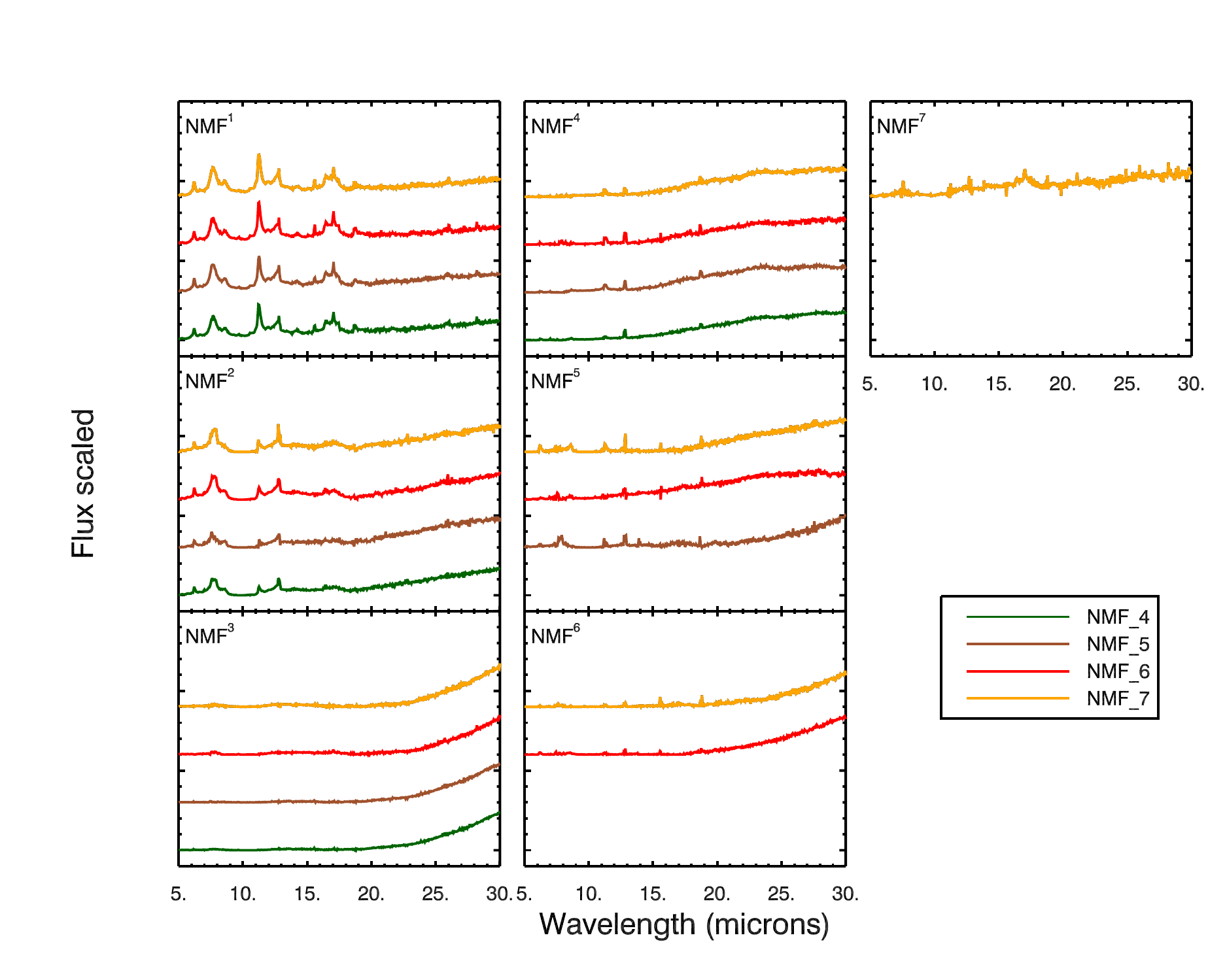}
\caption{The derived NMF spectral components for $NMF_{4}$-$NMF_{7}$, using only objects dominated by the third PAH component seen in Figure \ref{fig:template_plots}. Each NMF set is colour coded, with components ordered by similarity.}\label{fig:template_plots_PAH}
\end{figure*}

The third component captures the 6.2, 7.7, 8.6, 11.3, 12.7, 16.4 and 17.0 $\mathrm{\mu m}$ PAH features, and a cold dust slope at longer wavelength. There is also emission from Argon line [ArII] at 6.89 $\mathrm{\mu m}$ and Sulphur line [SIII] at 18.71 $\mathrm{\mu m}$. Its shape is similar to the \citep{Brandl:2006} average starburst template, based on 13 starburst galaxies. The ratio of the PAH features are very similar amongst component sets, but dust slope decreases with number of components. The reduction in dust slope for more complex NMF sets coincides with rising continuums seen in the fourth, sixth and seventh components.

The fifth component shows continuum emission up to 7 $\mathrm{\mu m}$ before dropping off at 10 $\mathrm{\mu m}$. It also shows strong emission from the Sulphur line [SIV] at 10.51 $\mathrm{\mu m}$. The remainder of the spectrum is noisy and featureless.

The eighth, ninth and tenth components show similarities to the first component. They show varying amounts of emission from the Neon lines, while the merged Oxygen and Iron lines appear as emission in the ninth component and absorption in the tenth. The variation of the first component in $NMF_{9}$ and $NMF_{10}$ compared to the other NMF sets is a result of the introduction of the ninth and tenth components and occurs because the NMF algorithm is using the freedom of extra components to break down the first into sub components.

\subsubsection{Physical interpretation of the components}

The first two components both show features associated with hot dust and gas emission and are likely to be related to AGN emission. The unified model of AGNs predict silicate emission from type 1 AGN and silicate absorption in type 2 AGN. More recently, the IRS spectra of type 2 quasi-stellar objects (QSOs) have shown silicate emission \citep{Sturm:2006}. \citep{Schweitzer:2008} have shown that the IRS spectra of 23 QSOs can be modelled with dusty narrow line region models, while \cite{Mason:2009} and \cite{Mor:2009} showed that clumpy torus models could also provide silicate emission for both type 1 and type 2 AGN. The fact we see a relatively stable silicate emission component amongst different NMF sets would suggest that silicate emission is occurring in more than just type 1 AGN and is a fundamental spectral component.

The third component is the main star formation component. It is dominated by PAH emission, often used as an indicator of star formation \citep[e.g.][]{Roussel:2001, Peeters:2004, Calzetti:2005, Kennicutt:2009}, and predominantly comes from photo-dissociation regions (PDRs) \citep{Roussel:2007, Peeters:2011}. For simpler NMF sets, the component also contains a rising continuum at longer wavelengths due to colder dust emission (T $\approx 50 K$), also associated with star formation \citep[e.g.][]{Calzetti:2007}. For the more complex NMF sets, the rising dust continuum is given its own component (e.g. the sixth and seventh). This indicates that although the colder dust and PAH emission both trace star formation, they come from different regions and the NMF algorithm uses the additional freedom of extra components to separate the two. We note that the PAH emission is extremely stable amongst all NMF sets and we do not see significant PAH emission in any other component. Previous studies show the ratio of PAH features vary with metallically and radiation hardness \citep[e.g.][]{Smith:2007}, yet we have one component with PAH emission.

To investigate the stability and lack of variation in the PAH emission features, we have re-run the NMF algorithm on objects from our original sample which are dominated by the third component. Figure \ref{fig:template_plots_PAH} shows the components from $NMF_{4}$ to $NMF_{7}$ for our reduced sample. The NMF algorithm now finds two components with PAH emission. The first shows emission at 6.2, 7.7, 8.6, 11.3, 12.7, 16.4 and 17.0 $\mathrm{\mu m}$, the second shows reduced emission for the 8.6, 11.3 and 12.7 $\mathrm{\mu m}$ PAH features and no emission at 16.4 and 17.0 $\mathrm{\mu m}$, while at longer wavelengths there is a rising continuum. The two new PAH components show a resemblance to those found in an NMF analysis of IRS spectro-imagery data for galactic PDRs \citep{Berne:2007}. Their first component, interpreted as emission from deep within the PDR, showed broad emission at 6.2, 7.8, and 11.4 $\mathrm{\mu m}$ and a rising continuum. The second component contained emission from the 6.2, 7.6, 8.6, 11.3, 12.7 and 17.4 $\mathrm{\mu m}$ PAH features, and was shown to be more dominant in regions closer to the star. 

\begin{figure*}
\includegraphics[width=18cm]{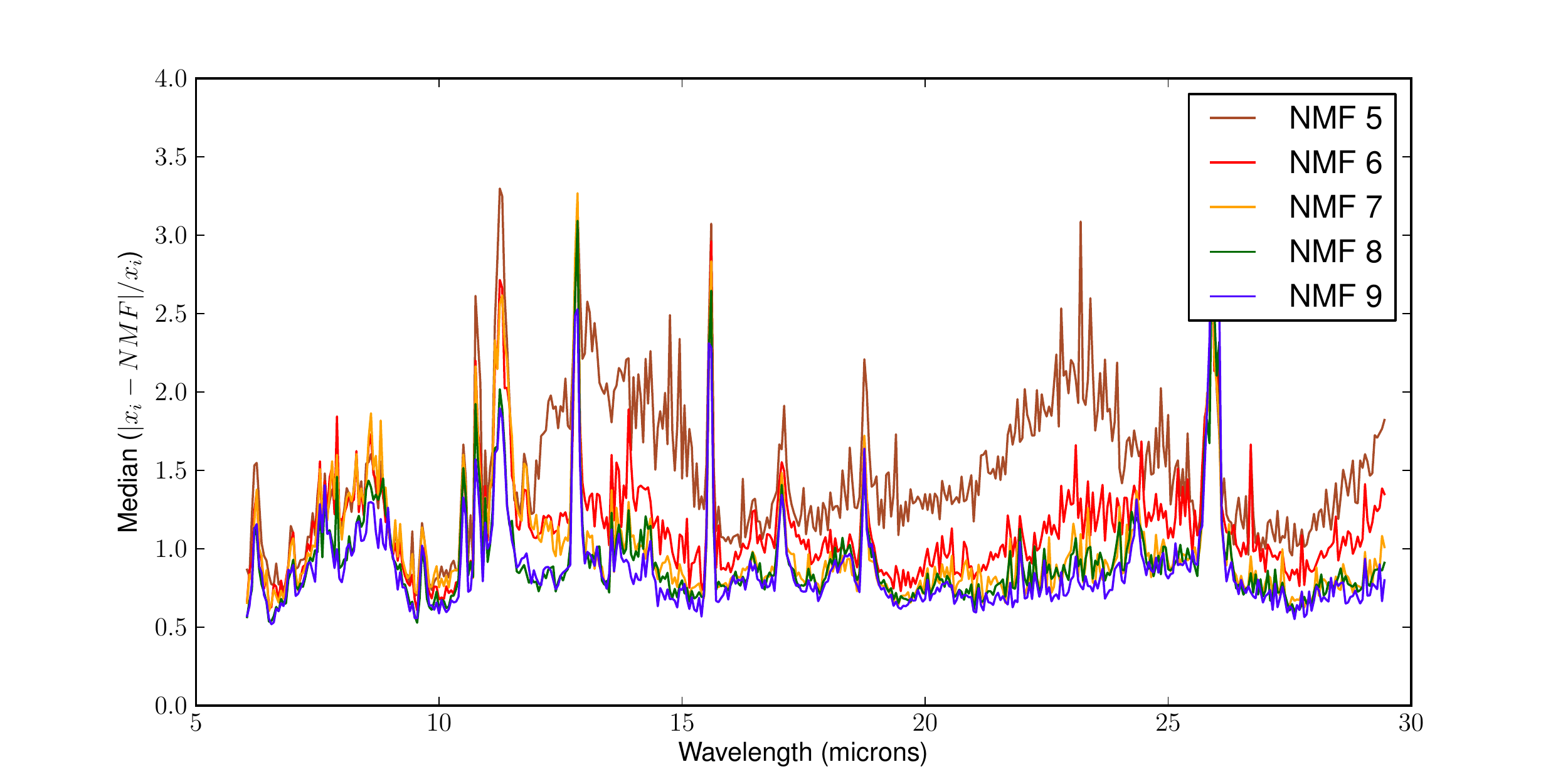}
\caption{The median absolute residuals, normalised by $\sigma$, for NMF sets $NMF_{5}$-$NMF_{9}$. The residuals show all NMF sets fail to capture the variance in many of the emission lines. However, for NMF sets $NMF_{7}$ and above, the residuals for the underlying continuum are down to 1 $\sigma$.}\label{combined_res}
\end{figure*}

By restricting the sample to objects dominated by star formation, the NMF algorithm does not need to use components to separate out hotter dust from AGN, and uses the additional freedom to separate out the PAH emission. The PAH emission in our original third component is therefore capturing the average PAH emission from galaxies. 

Components four, six and seven from Figure \ref{fig:template_plots}, all contain rising continuums, though with varying slopes and is capturing dust emission at different temperatures. The fact we see numerous components with varying slopes suggests that the colder greybody emission of dust varies considerably amongst galaxies. The seventh component also contains a bump at around 8 and 12 $\mathrm{\mu m}$. The bumps help build up a silicate absorption feature at 10 $\mathrm{\mu m}$, this component is therefore important for dusty galaxies.

\begin{figure}
\includegraphics[width=9cm]{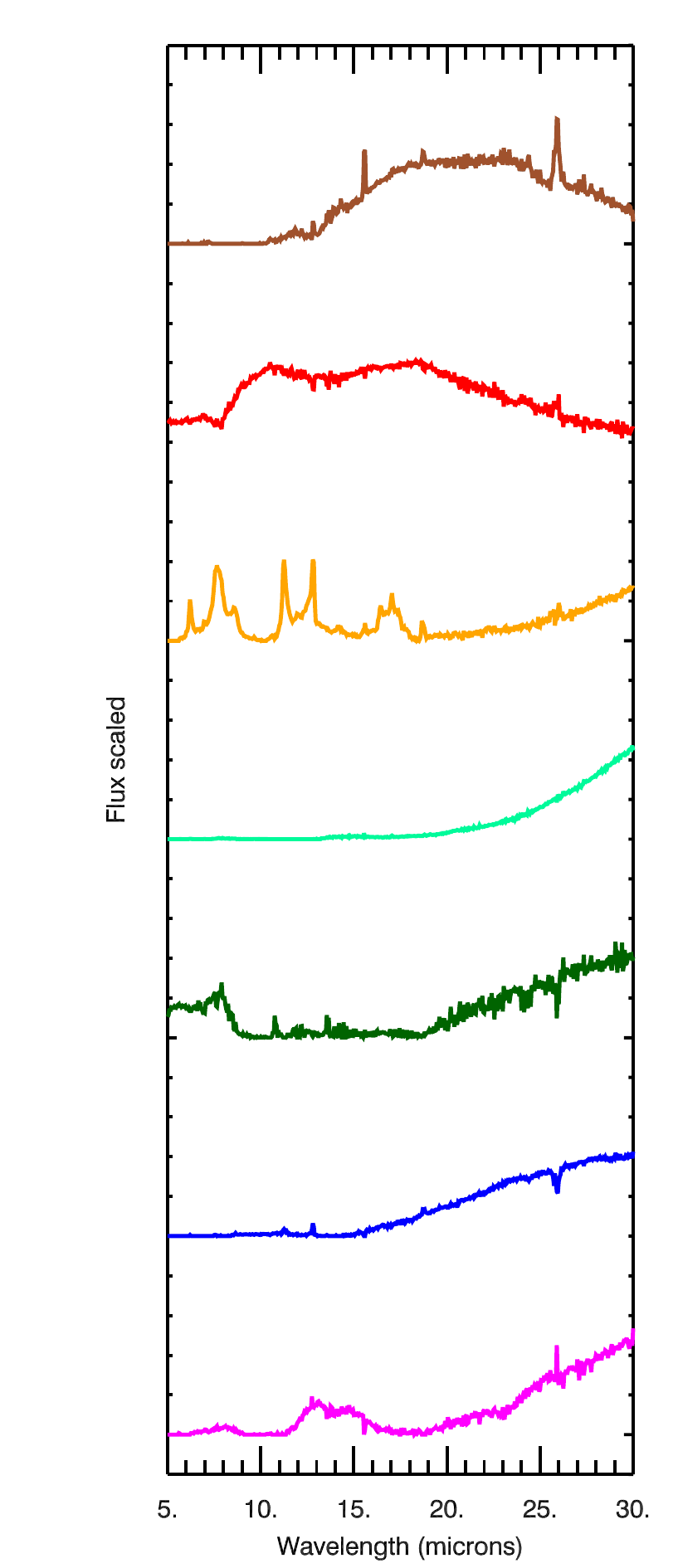}
\caption{The 7 components from $NMF_{7}$, corresponding to the yellow components in Figure \ref{fig:template_plots}. The new colour coding is used to identify the different components in subsequent figures.}\label{7_components}
\end{figure}

To further investigate the components, we can begin to look at how they contribute to different types of spectra. In order to simplify the analysis and to provide a simple set of components, we restrict our components to those from $NMF_{7}$. Our choice of seven is more qualitative than quantitative, as we have already shown that a quantitative analysis requires more than 14 components. To validate our choice, we have studied the median, absolute residuals of NMF fits to the CASSIS sample with $NMF_{5}$ to $NMF_{9}$, shown in Figure \ref{combined_res}. The residuals are high for some of the emission lines, particularly the PAH features, because our components capture the average line emission. However we note that by seven components, the residuals for the underlying continuum are down to 1 $\sigma$ and there is little advantage in using more complicated sets. By choosing seven, we believe we strike the balance between having enough simplicity to have a useful and physically intuitive NMF set of components, whilst being able to reconstruct the general spectral shape. The seven components are re-plotted in Figure \ref{7_components}. 

\subsubsection{$NMF_{7}$ fits to example galaxy spectra}
We now examine the NMF fits to spectra of different types of galaxies in order to show how contributions from components vary and that our $NMF_{7}$ set can capture the general shape of different types of spectra. Our example fits, along with the corresponding residuals (i.e. data-fit) can be found in Figures \ref{example_gal_fits} and \ref{example_gal_fits2}. The first plot in Figure \ref{example_gal_fits} shows the NMF fit to the Blue Compact Dwarf (BCD) KUG 1013+381, observed as part of the IRS Guaranteed Time Observation (GTO) program. BCDs tend to be small galaxies with low metallicity, that have undergone a recent burst of star formation but have suppressed star formation compared to typical starburst galaxies \citep{Wu:2006}. 

Our NMF fit shows component one makes a significant contribution, suggesting there is some hot dust. Component four also makes a large contribution, indicating emission from colder dust. Components six and seven, both containing dust slopes at longer wavelengths, also contribute. There is very little emission from component two, which we believe is associated with the inner surface of an AGN and there is very little emission from the third 'PAH' component. The residual plot shows the $NMF_{7}$ set can construct the underlying continuum, however the [SIV], [NeIII] and [SIII] emission lines are underestimated.

Our second NMF fit is to the ULIRG and type 1 Seyfert galaxy, Markarian 231. Unlike, KUG 1013+381, the second 'silicate emission' component makes a contribution, and the other, warmer dust components such as six and seven contribute as much power to the longer wavelengths as the fourth component. There is very little contribution from the third component. Residuals show the fit is reasonable except beyond 25 $\mathrm{\mu m}$, where there appears to be some instrumental artefact in the spectra.

The third fit is to PG 1211+143, also a type 1 Seyfert galaxy. The second component dominates the emission of this object. The first, fifth and sixth component make comparable contributions. The residual plot shows that our $NMF_{7}$ set slightly over estimates emission from the [NeIII] and [OIV] lines. 

The fit to the ULIRG and type 2 Seyfert galaxy, Markarian 273, is dominated by emission from the fourth 'cold dust' component. Residuals show the NMF components underestimate some of the emission lines, particularly the [NeIII] line. The continuum appears to be well reconstructed by the NMF components.

Our final two fits in Figure \ref{example_gal_fits} are to the starburst galaxies, NGC 3301 and NGC 3256. The third component contributes in the shorter wavelengths, while the colder dust components,  four and six, contribute at longer wavelengths. The residuals show the components are capable of reconstructing the continuum, but fail to capture the emission lines accurately.

Four additional example fits are shown in Figure \ref{example_gal_fits2}. The first is to LINER, 3C270. The first, second and fifth components are the main contributors. while the residuals show the fit can reconstruct the continuum, but underestimate the 12.8 $\mathrm{\mu m}$ Neon line. The submillimeter galaxy GN26 is over a short wavelength region and the spectrum is quite noisy. Our final two fits are to quasar PG0804+761 and ULIRG IRAS 10378+1108. As with other type 1 AGN, the second component dominates emission. Our $NMF_{7}$ set fails to model the full width of the very broad silicate emission feature at 9.7 $\mathrm{\mu m}$, however the rest of the continuum is well reconstructed. Our NMF fit to the ULIRG IRAS 10378+1108 dominates the emission, while the residuals show the NMF set slightly overestimate the greybody emission longwards of 27 $\mathrm{\mu m}$.

In addition to galaxy spectra, we also fit our $NMF_{7}$ set to the average spectral templates from the IRS spectral ATLAS of galaxies \citep{Hernan-Caballero:2011}. Table \ref{tabla_templates} in Appendix C gives more details on the sources used for the ATLAS average templates. As can be seen in Figure \ref{example_av_fits}, the change in contributions for different types of object is consistent with those in Figure \ref{example_gal_fits}. The continuum is well constructed for all average templates, however the residuals show the emission lines are not accurately reconstructed, especially for the average LINER template.

Overall, our fits show for Seyfert galaxies, the first and second component, along with the warmer dust components of five, six and seven are all important, though their contributions vary. For the starburst galaxies, the third and fourth component play a more important role. The residual plots show that our NMF set is capable of reconstructing the continuum to a reasonable accuracy, however some of the emission lines are not always fitted well. This is to be expected since, as we have previously shown, the components capture the 'average emission' of spectral lines. To accurately fit continuum and emission lines, our Bayesian evidence calculation has shown we would need an NMF set with more than 14 components. The goal of this paper is to find a physically intuitive component set, which requires a balance between number of components and ability to reconstruct spectra. We believe Figure \ref{combined_res} and \ref{example_gal_fits} shows our $NMF_{7}$ set fits this requirement.

\begin{figure*}
\includegraphics[width=18cm]{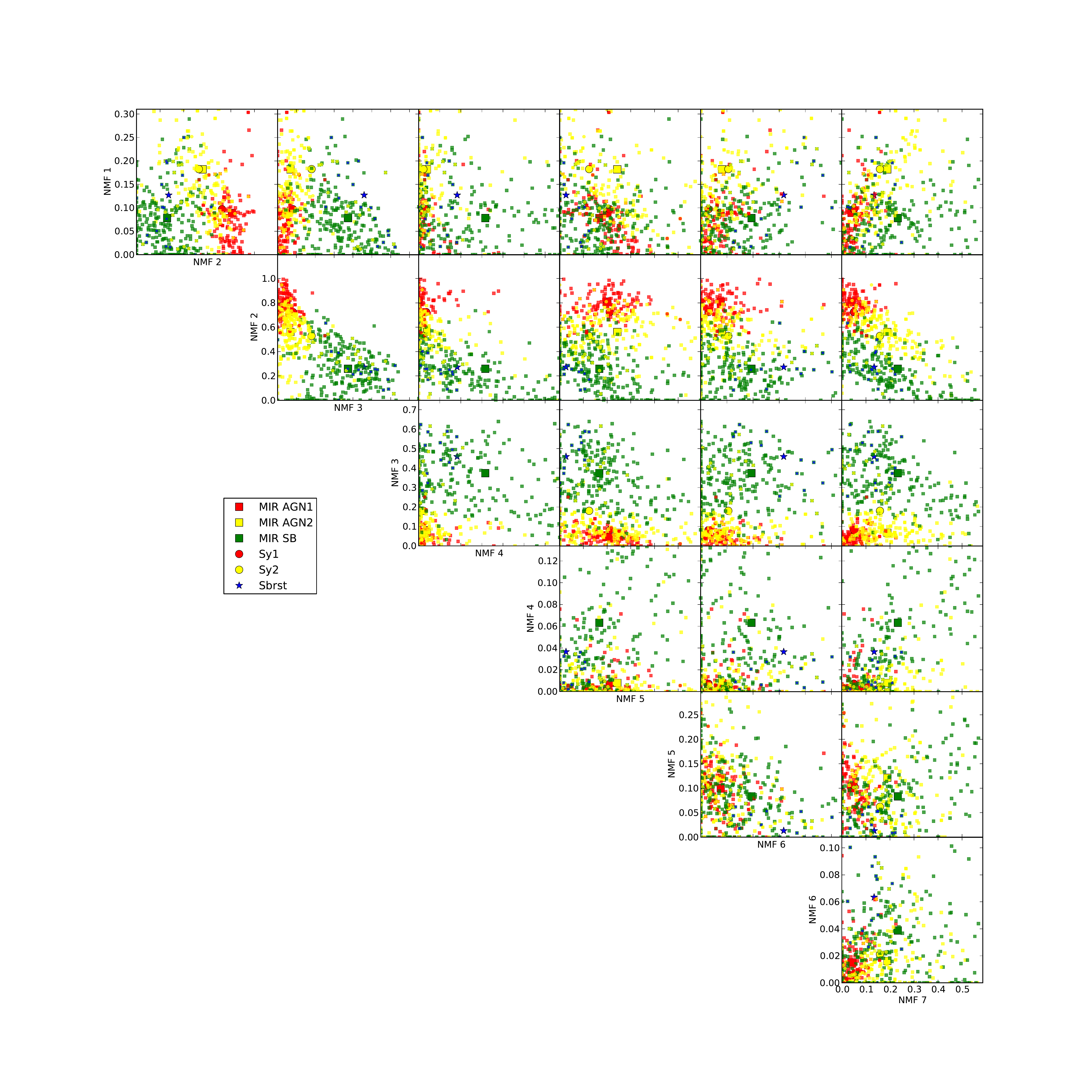}
\caption{The distribution of objects/spectra from the ATLAS groups: MIR AGN1, MIR AGN2, MIR SB, Sbrst, Sy1 and Sy2 in our 7D space defined by the $NMF_{7}$ set. Symbols and colours for the different groups are described in the legend. The position of the average template for each group is marked by a larger symbol.}\label{Av_temps_in_7DNMF}
\end{figure*}

To illustrate how the components contribute to a number objects, we can use the weightings provided by the NMF fits as multidimensional co-ordinates. Each galaxy is now a point in a seven dimensional space we call NMF space. We use classifications from the IRS spectral ATLAS of galaxies \citep{Hernan-Caballero:2011} to investigate what regions of NMF space are associated with different types of galaxies. The ATLAS collection contains spectra from a number of observing programs. They provide optical classifications from the literature and three additional MIR classifications: MIR SB, MIR AGN1, MIR AGN2 based on the fractional contribution from a PDR component used during spectral decomposition. The AGN subgroups MIR AGN1 and MIR AGN2 are subsets of AGN, classified by whether spectra show silicate emission or silicate absorption. Figure \ref{Av_temps_in_7DNMF} shows how objects from the ATLAS groups: MIR AGN1, MIR AGN2, MIR SB, Sbrst, Sy1 and Sy2 are distributed in the seven dimensional NMF space. 

As can be seen in Figure \ref{Av_temps_in_7DNMF}, the Seyfert 1 and MIR AGN1 objects all lie in a region with low contribution from $NMF^{1}_{7}$, high contribution from $NMF^{2}_{7}$ and very little contribution from $NMF^{3}_{7}$. The Seyfert 2 and MIR AGN2 objects are found in a region with a higher contribution in $NMF^{1}_{7}$, less or very little contribution from $NMF^{2}_{7}$ and very little contribution from $NMF^{3}_{7}$. Starburst like objects on the other hand require little contribution from either $NMF^{1}_{7}$ or $NMF^{2}_{7}$, and a high contribution from $NMF^{3}_{7}$. 

We note that the components most influential in separating out the different objects are the components one, two and three. Less influential but still significant are the colder dust components $NMF^{4}_{7}$ and $NMF^{6}_{7}$. They contribute very little to objects classified as AGN, while the contribution for starbursts show a large variation. This fits in with our earlier interpretation that these two components represent obscured star formation components which vary more than the PAH features seen in $NMF^{3}_{7}$. The remaining two components are the least significant. There is a slight difference in contribution between AGN 1 objects and the other two classes, while $NMF^{7}_{7}$ separates out type 1 and type 2 objects to a certain extent.  

\subsection{Gaussian Mixtures Modelling}\label{subsec:GMM}
We have shown NMF space is capable of separating out different types of objects. We now model how objects separate out in this multidimensional space by applying the parametric technique Gaussian mixtures modelling (GMM). GMM has already been successfully applied to the colour and redshift space of galaxies \citep{Davoodi:2006}. GMM assumes the distribution of objects can be modelled by a series of clusters, each described by a multidimensional Gaussian. We use the GMM software from the Auton Lab \footnote{http://www.autonlab.org}\citep{Moore:1999} to model the distribution of the CASSIS sample in our 7 dimensional NMF space. The software uses the Expectation Maximisation algorithm to learn the position and size of the clusters and uses the Akaike Information Criterion (AIC;\cite{Akaike:1974}) to select how many are needed to describe the distribution of objects. 

We find that 8 clusters are required to adequately model the distribution. Each cluster describes a probability density function (PDF) for any position in NMF space. By using an objects position in NMF space, we can assign it to one of the 8 clusters.\footnote{Every position in NMF space has eight PDF values associated with it (one for each cluster). Using the highest probability density provides the optimal (maximum likelihood) classification. However, since the PDFs overlap, this will not provide the best classification for the population statistics. We therefore take the same approach as \cite {Davoodi:2006} and randomly assign each galaxy to a cluster, with probability proportional to the PDF values at the galaxies position in NMF space.}Table \ref{GMMtable} shows how some of the ATLAS classified sources are distributed across the 8 clusters, with clusters ordered by their normalisation (i.e. how many objects are in that cluster). As can be see in Table \ref{GMMtable}, the majority of objects are contained within the first five clusters. The normalisations associated with the remaining clusters (i.e. how many objects they capture) are also very small. We therefore use the first five clusters to define a classification scheme.

\begin{figure*}
\includegraphics[width=18cm]{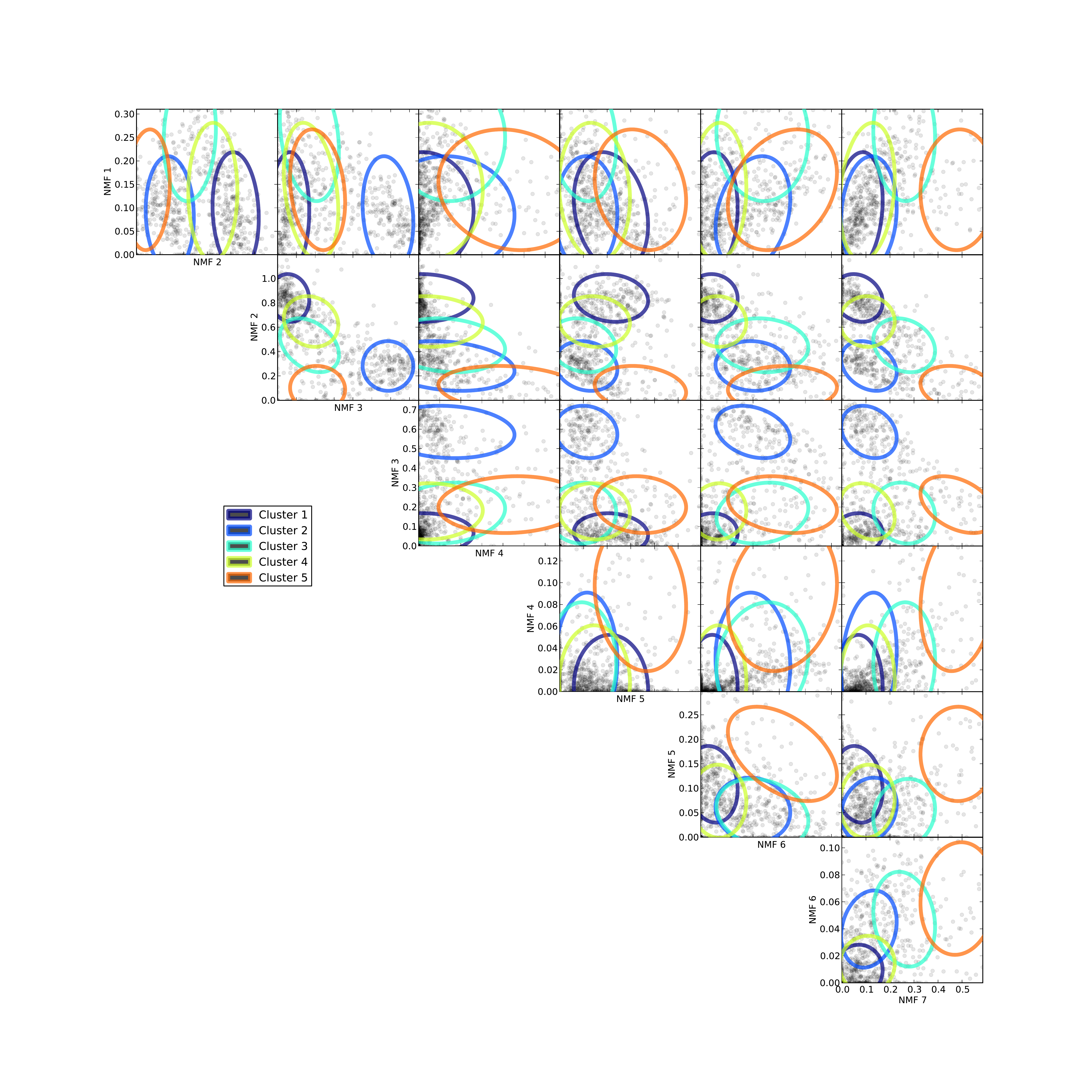}
\caption{NMF space for 7 templates. CASSIS objects used for NMF and GMM are also plotted. The ellipses represent the different clusters found through Gaussian Mixtures Modelling}\label{GMM_CASSIS_7Dplot}
\end{figure*}

The location in NMF space of the first five clusters can be seen in Figure \ref{GMM_CASSIS_7Dplot}. Each cluster is represented by its 1 sigma contour. The CASSIS sample used for training the Gaussian Mixtures modelling are also plotted.

As can be seen in Figure \ref{GMM_CASSIS_7Dplot} and classifications in Table \ref{GMMtable}, cluster one captures nearly all the Seyfert one galaxies, and some Seyfert two galaxies. Cluster two contains a significant number of objects previously classified as starbursts, while cluster three contains a large proportion of the remaining Seyfert two objects. The position of cluster four indicates this could be an intermediary group between typical Type one and Type two galaxies. The fifth cluster contains just over a fifth of those objects classified as starbursts in the MIR and no optically classified starbursts. Its position in NMF space also suggests it captures those objects which are dusty starbursts.

We conclude that cluster one is related to Seyfert 1 galaxies, cluster two with starbursts, cluster three with Seyfert two galaxies and cluster four for galaxies showing signs of both Seyfert one and Seyfert two (e.g. Type 1.5). The fifth cluster captures those galaxies which are dusty and obscured. The clusters can be used as a classification scheme by taking any IRS galaxy spectrum, fitting with $NMF_{7}$ set and using the corresponding weights to identify what cluster the object is associated with. 

\begin{figure}
\includegraphics[width=8.5cm]{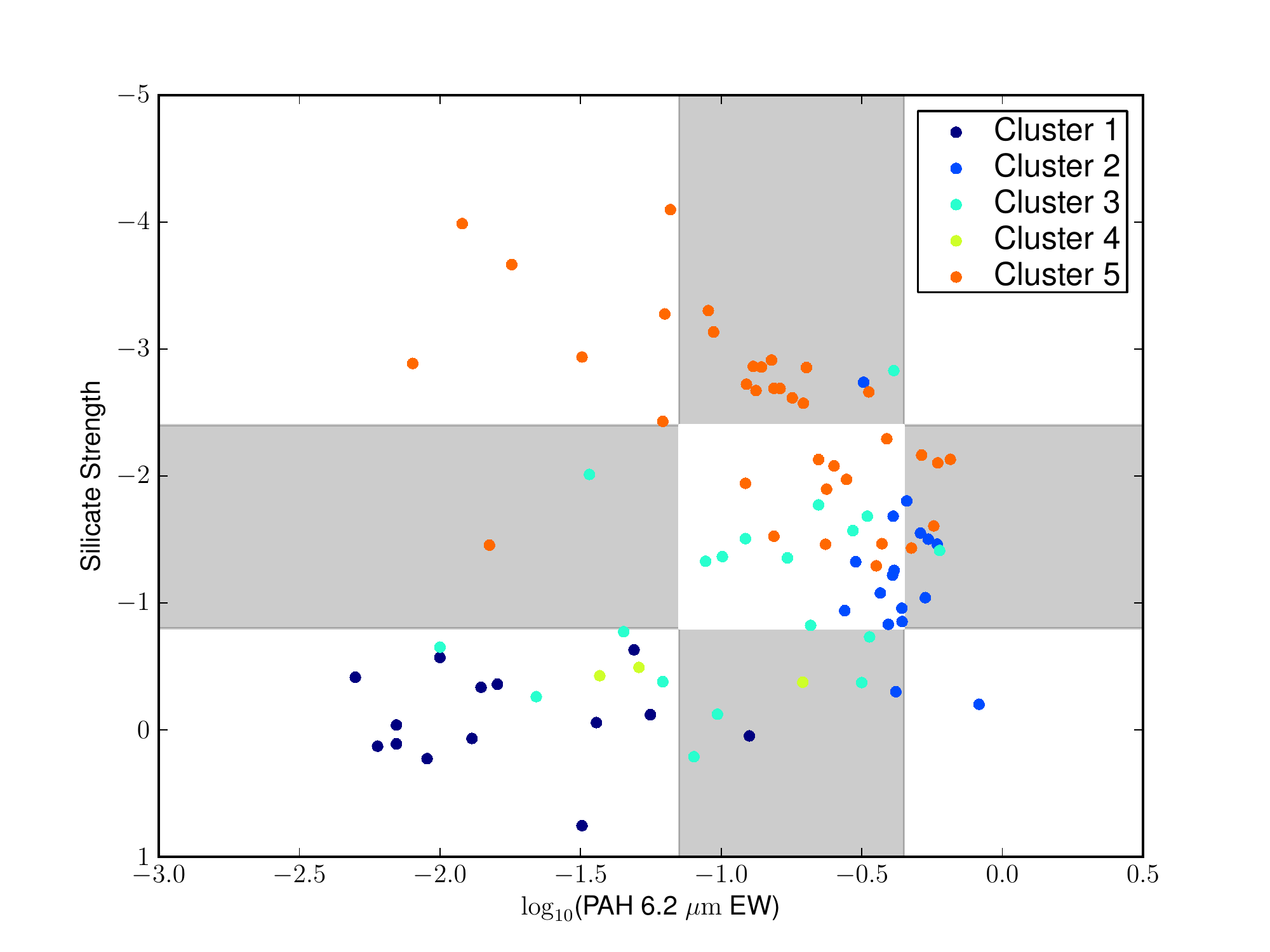}
\caption{The \protect\cite{Spoon:2007uq} diagram showing Silicate strength versus the 6.2 $\mathrm{\mu m}$ PAH equivalent width. The plot is separated into the different Spoon classes and objects are colour coded by our GMM classification.}\label{SpoonGMM}

\end{figure}

We compare our classification scheme to the \cite{Spoon:2007uq} diagram, which classified ULIRGs via the strength of their $9.7 \mathrm{\mu m}$ silicate feature and 6.2 $\mathrm{\mu m}$ equivalent width. Figure \ref{SpoonGMM} shows 89 ULIRGs in the \cite{Spoon:2007uq} diagram, colour coded by our our classification. Seyfert one classified galaxies lie on the far left of the bottom horizontal branch, corresponding to a 1A and 1B Spoon classification, Seyfert two classified galaxies span the horizontal branch and 2B Spoon classification. The starburst classified objects are located in the far bottom right of the Spoon diagram, while dusty objects are spread out across the diagonal branch. Only three objects are classified as Type 1.5 and they lie on the horizontal branch, in-between the Seyfert one and Seyfert two classified galaxies. 

Comparing the success rates of different classification schemes, without knowing the 'true' classification is always problematic, however our classification scheme is consistent with the \cite{Spoon:2007uq} interpretation of Figure \ref{SpoonGMM} in terms of the location of starbursts, AGN dominated objects and dusty objects. Unlike the Spoon diagram, our classification scheme can also distinguish between Seyfert one and Seyfert two galaxies.

We have shown our classification scheme is just as successful as the Spoon classification. However, our classification has three distinct advantages over \cite{Spoon:2007uq}. First, \cite{Spoon:2007uq} only use the $9.7 \mathrm{\mu m}$ silicate feature and 6.2 $\mathrm{\mu m}$ PAH equivalent width to separate out classes. By using the NMF components as a basis for our GMM based classification scheme, we make use of the whole MIR region to classify objects. This also enables us to classify objects where the $9.7 \mathrm{\mu m}$ silicate feature and 6.2 $\mathrm{\mu m}$ PAH equivalent width are not available or difficult to measure. Secondly, our classification scheme is modelled on the number density of our CASSIS sample in NMF space. Since our sample contains a large variety of objects, any sample biases will have a small affect on the outcome of our classification scheme. The Spoon classes on the other hand, are chosen based on arbitrary cuts in the $9.7 \mathrm{\mu m}$ silicate feature and 6.2 $\mathrm{\mu m}$ PAH equivalent width. Thirdly, because our clusters describe a probability density function, we can give an indication of how likely a galaxy could be found in any one of the five clusters. For example, in Table \ref{GMMexample} we show the probability of being in any of the five clusters for some famous objects.

We make our classification tool publicly available on the arxiv and at \url{https://github.com/pdh21/NMF_software/}.

\begin{table*}
\input{GMM_table_7_1000s_0p1eps_v4}

\caption{The percentage of ATLAS classified objects in each cluster for 7 NMF templates. The first column indicates the cluster number. The second column shows the probability that a CASSIS object is in that cluster (i.e. how many objects can be found in it). The remaining columns contain the percentage of ATLAS classification in each cluster.}\label{GMMtable}
\end{table*}

\begin{table*}
\begin{tabular}{|c|c|c|c|c|c|}
\hline
& Cluster1& Cluster2& Cluster3& Cluster4& Cluster5\\
Object & Sy1 & Sbrst & Sy2 & Sy1.5 & Dusty SB \\ \hline
Arp220 & 0.00 & 0.23 & 0.41 & 0.02 & 0.34 \\
Mrk231 & 0.32 & 0.00 & 0.34 & 0.32 & 0.02 \\
PG1211+143 & 0.92 & 0.00 & 0.00 & 0.08 & 0.00 \\
IRAS10565+2448 & 0.00  & 0.71 & 0.25 & 0.00 & 0.04 \\
IRAS10378+1109 & 0.00 & 0.01 & 0.06 & 0.00 & 0.93 \\
\hline
\end{tabular}
\caption{The approximate probability of being in one of the five clusters in our GMM based classification scheme.}\label{GMMexample}
\end{table*}

\subsection{SF-AGN contribution}

We have shown that the NMF components are capable of distinguishing between the objects showing extreme star formation or AGN activity. We now use them to introduce a diagnostic to quantify the contribution from star formation and AGN. Unlike other diagnostics, ours employ the whole MIR spectrum to disentangle the SF versus AGN contributions, and it is not based on specific features for which we need to know information on their origin.

For AGN, $NMF^{1}_{7}$ and $NMF^{2}_{7}$ are the most important and bear the physical features we know to originate from AGN tori. We therefore adopt $NMF^{1}_{7}$ and $NMF^{2}_{7}$ as contribution from AGN. For star formation, the third component is the most important, however we argue that the fourth and fifth components are also required as they contain the colder dust associated with obscured star formation. This is especially important for objects like Arp 220 which are known to be predominantly powered by star formation but have less than average PAH emission compared to other submillimeter galaxies \citep{Pope:2008}. We do not include $NMF^{6}_{7}$ and $NMF^{7}_{7}$ in our diagnostic. These components contribute to both AGN and starbursts and we have interpreted them as arbitrary dust components that are not specifically associated with star formation or AGN activity. Our diagnostic is taken as the ratio of MIR luminosity from the following components:
\begin{equation}
\frac{\mathrm{star formation}}{\mathrm{AGN}}=\frac{L_{NMF3} + L_{NMF4} + L_{NMF6}}{L_{NMF1} + L_{NMF2}}
\end{equation}

\subsubsection{Comparison to other MIR diagnostics}
We now show this diagnostic compared to other MIR diagnostic plots quantifying star formation and AGN contribution. 

\cite{Farrah:2009} applied Bayesian inferencing and graph theory to a data set of 102 mid-infrared spectra. By examining how position in the network was related to other parameters  (e.g.infrared luminosity, optical spectral type and black hole mass) they concluded that the network depicted the evolutionary scheme of ULIRGs, with different branches relating to Starburst+AGN and luminous AGN.

\begin{figure}
\begin{tabular}{c}
\includegraphics[width=7.5cm]{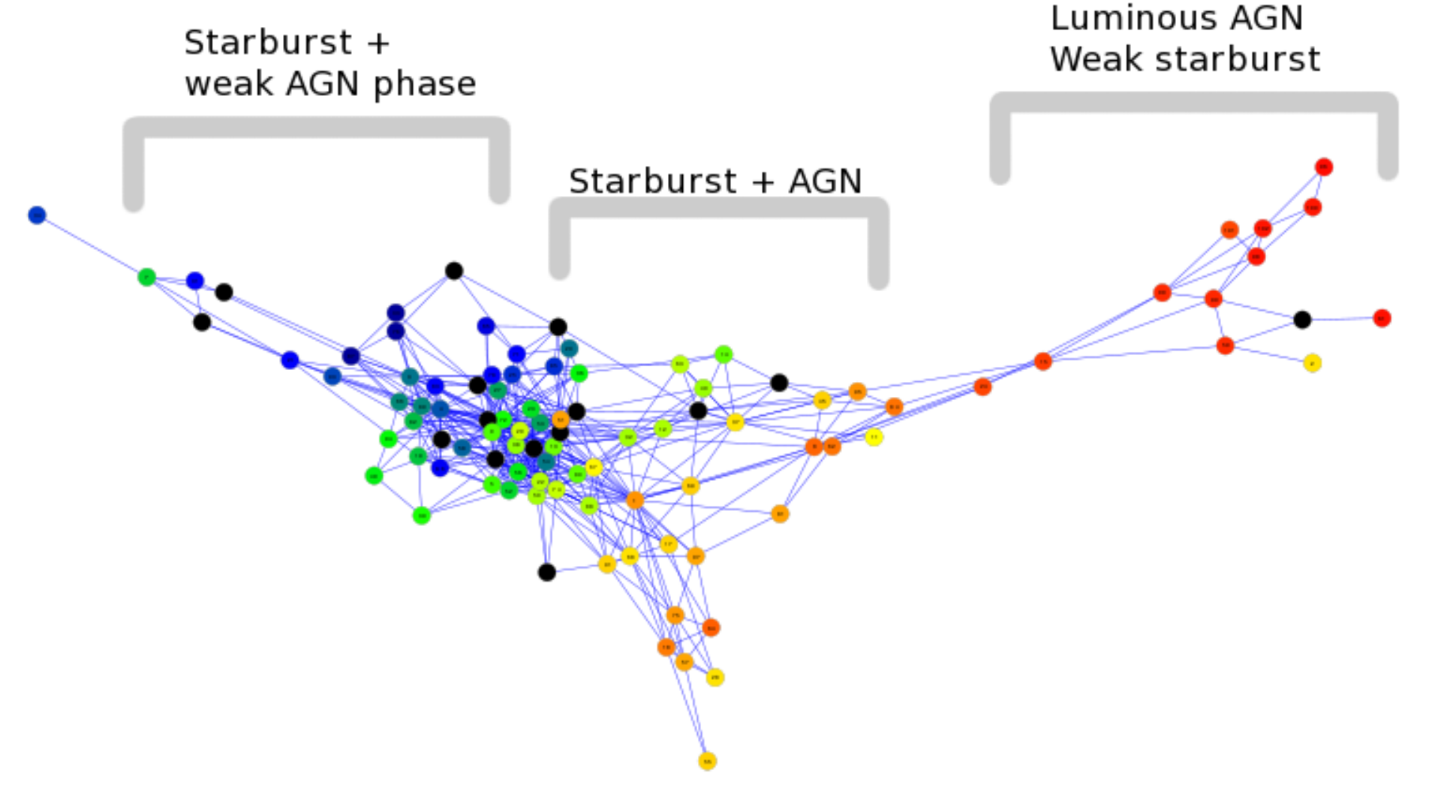} \\
\includegraphics[width=7.5cm]{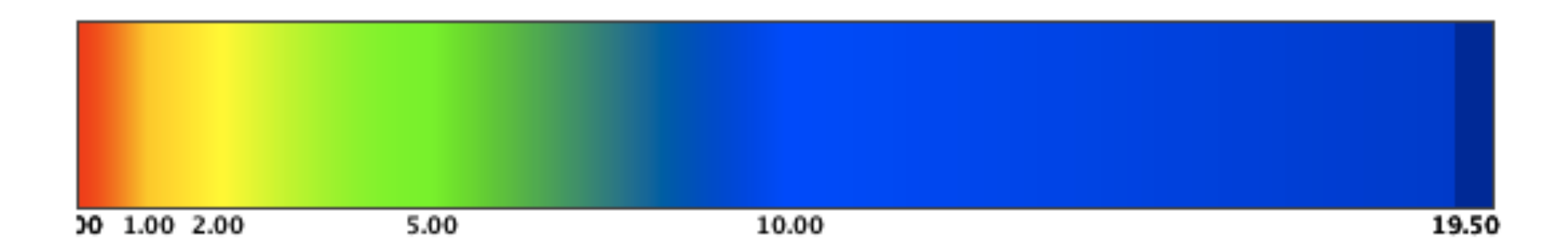} \\
\end{tabular}
\caption{The network diagram along with interpretation from \protect\cite{Farrah:2009}. Starbursts dominate the left hand side of the network. As the AGN becomes more dominant, galaxies move to the right and finally on to one of the two branches. The Nodes are colour coded by our NMF diagnostic. Nodes in black are where spectra are not available.}\label{figevo_plots}
\end{figure}

We now investigate how our $NMF_{7}$ set relate to the same network by decomposing the \cite{Farrah:2009} sample with our NMF components and colour-coding the network by our NMF diagnostic. The connections are taken from \cite{Farrah:2009} and we use the same Cytoscape software \footnote{Available from http://cytoscape.org/.} to produce the network. We note that our network is not identical to that in \cite{Farrah:2009} due to the random seed starting position used by the spring-embedded algorithm in Cytoscape. The two main branches seen in \cite{Farrah:2009} are still seen in Figure \ref{figevo_plots}, with the lower and right hand branches corresponding to the Starburst+ AGN and Luminous AGN branches respectively. Each galaxy is colour coded by our new NMF diagnostic.

As can be seen in Figure \ref{figevo_plots}, our NMF diagnostic is consistent with the interpretation that star formation occurs on the left hand side of the network, with AGN activity increasing as we move to the right. The right hand branch appears to be AGN dominated, as was concluded in \cite{Farrah:2009}. 
 
\begin{figure}
\includegraphics[width=8.5cm]{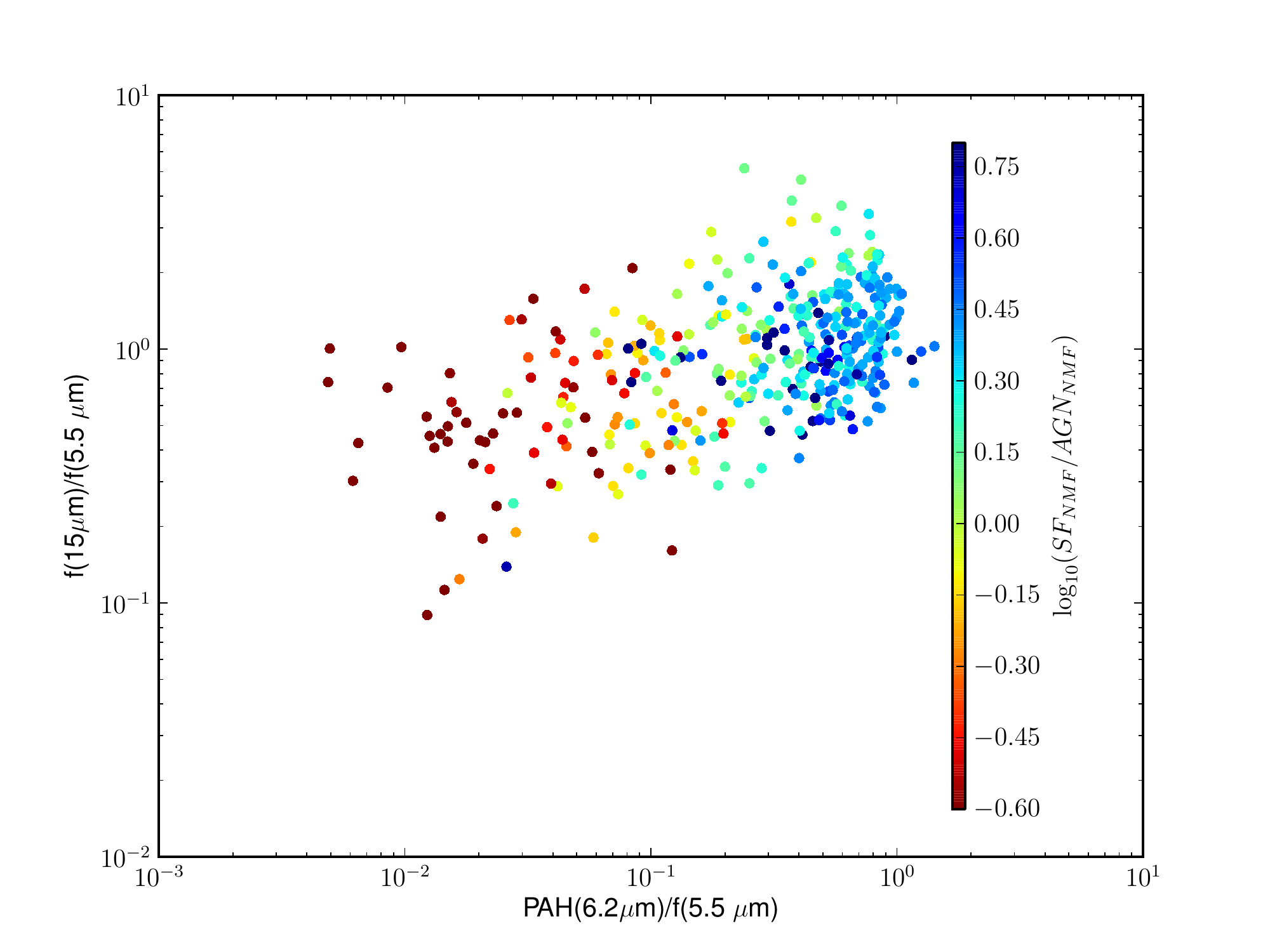}
\caption{The ratio of 15 to 5 $\mathrm{\mu m}$ continuum flux, against the $6.2 \mathrm{\mu m}$ PAH flux to 5 $\mathrm{\mu m}$ continuum flux, as seen in \protect\cite{Armus:2007}. Points are colour coded by our NMF diagnostic.}\label{cont_diag}
\end{figure}

Our second comparison is with the diagnostic diagram introduced by \cite{Laurent:2000} and modified for Spitzer by \cite{Armus:2007}. The diagrams use the integrated continuum flux from $14-15 \mathrm{\mu m}$, the integrated continuum flux from $5.3-5.5$ and the $6.2 \mathrm{\mu m}$ PAH flux to indicate fractional contributions from AGN and starbursts. Figure \ref{cont_diag} shows the same diagnostic plot, plotted with objects from the CASSIS database with measurements of the continuum and $6.2 \mathrm{\mu m}$ flux taken from the CASSIS database. The points are colour coded by our NMF diagnostic.

Objects with a high NMF SF-AGN ratio are located in the top right while objects with a low NMF SF-AGN ratio lie in the bottom left. This is consistent with the simple linear mixing lines indicating AGN and star formation fraction seen in \cite{Armus:2007} and \cite{Petric:2010}.

\begin{figure}
\includegraphics[width=8.5cm]{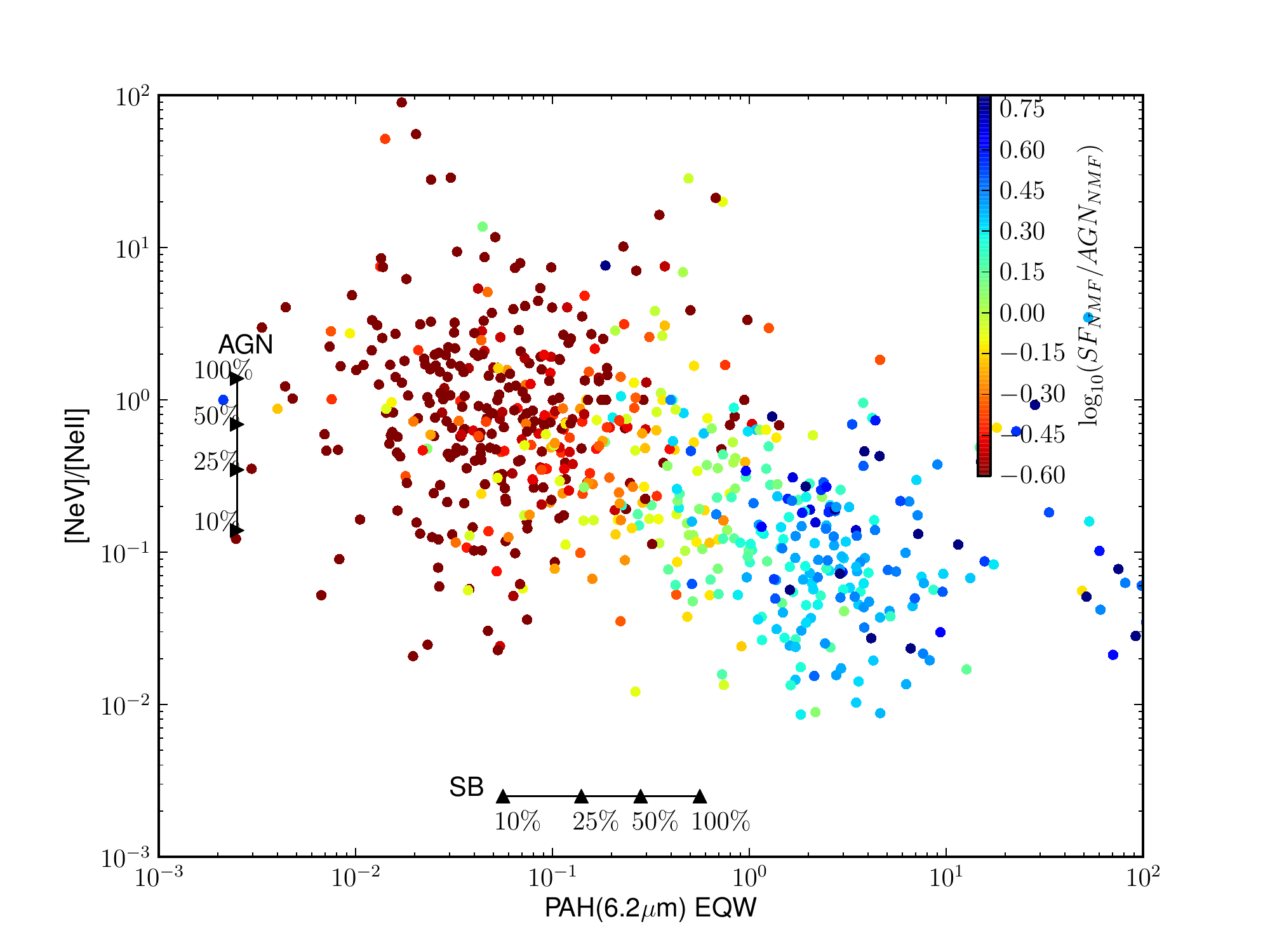}
\caption{The [NeV]/[NeII] ratio vs the PAH 6.2 $\mathrm{\mu m}$ equivalent width. The points are those objects in the CASSIS database that have a redshift and an estimate for the three lines. The points are colour coded by our NMF diagnostic. We also show the 100\%, 50\%, 25\%, and 10\% AGN and starburst linear mixing contributions taken from \protect\cite{Armus:2007}}.\label{Ne_diag}
\end{figure}

Our third and fourth comparison is with diagnostic diagrams using emission lines. We plot all spectra in the CASSIS database that have a known redshift and measurable emission line. Line measurements are made with the PAHfit software \citep{Smith:2007}. Figure \ref{Ne_diag} shows the ratio of Neon forbidden lines [NeV] and [NeII] against the PAH 6.2 $\mathrm{\mu m}$ equivalent width, colour coded by the NMF diagnostic. We indicate the fractional AGN and starburst contribution to the MIR luminosity from the [NeV]/[NeII] (vertical) and 6.2 $\mathrm{\mu m}$ PAH EQW (horizontal) assuming a simple linear mixing model. In each case, the 100\%, 50\%, 25\%, and 10\% levels are marked. The 100\% level is set by the average detected values for the [NeV]/[NeII] and PAH 6.2 $\mathrm{\mu m}$ equivalent width among AGN and starbursts respectively, as discussed in \cite{Armus:2007}.

We see that our diagnostic is consistent with star formation dominated objects being located in the bottom right of the plot, while objects with higher AGN contribution are located in the top left. 
\begin{figure}
\includegraphics[width=8.5cm]{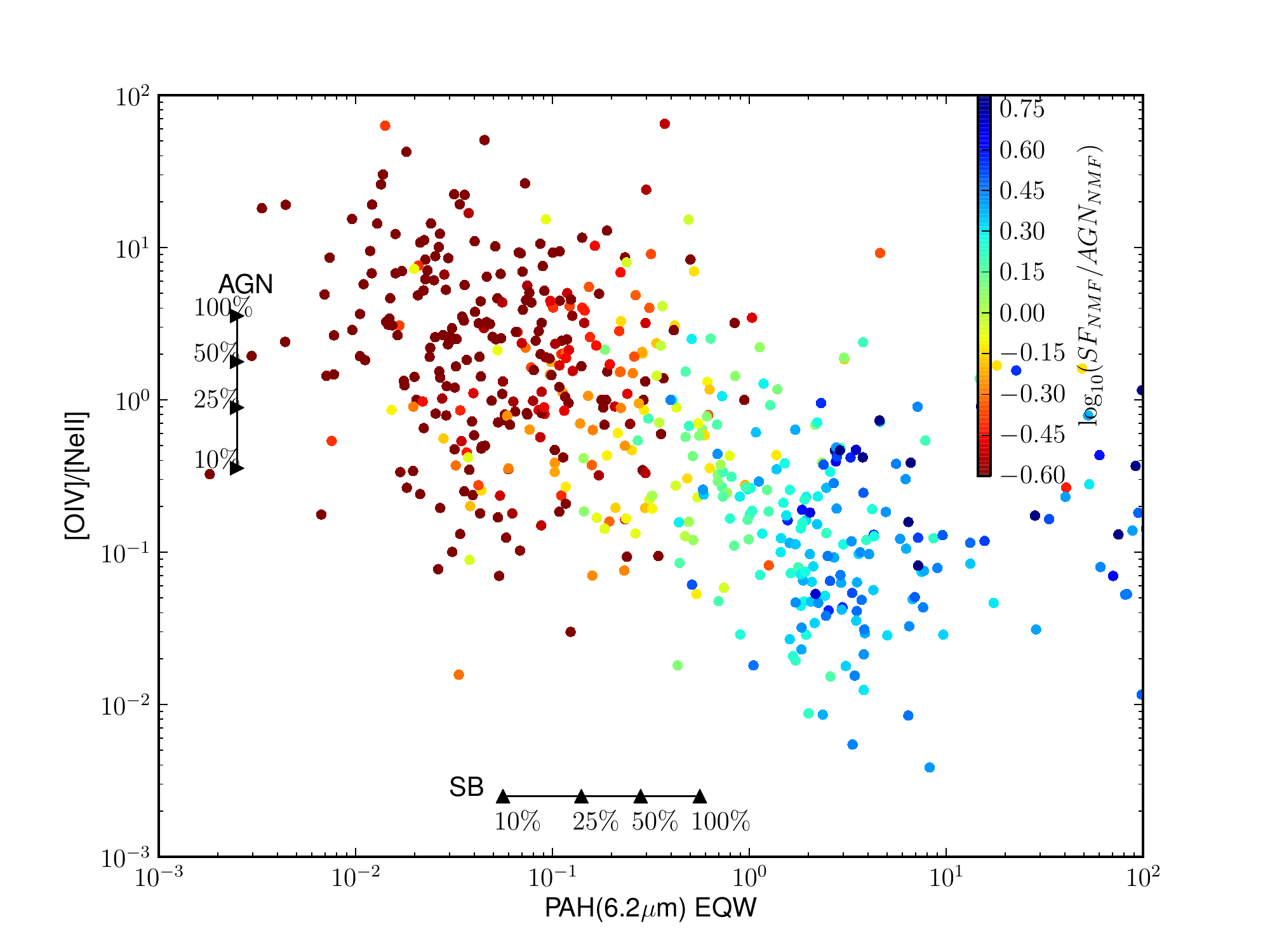}
\caption{The [OIV]/[NeII] ratio vs the PAH 6.2 $\mathrm{\mu m}$ equivalent width. The points are those objects in the CASSIS database that have a redshift and an estimate for the three lines. The points are colour coded by our NMF diagnostic. We also show the 100\%, 50\%, 25\%, and 10\% AGN and starburst linear mixing contributions taken from \protect\cite{Armus:2007}}\label{ox_diag}
\end{figure}

The third diagnostic diagram uses the [OIV] and [NeII] ratio vs PAH 6.2 $\mathrm{\mu m}$ equivalent width. As in Figure \ref{Ne_diag}, we colour code the points by NMF diagnostic and indicate the fractional AGN and starburst contributions as discussed in \cite{Armus:2007}. Our plot can be seen in Figure \ref{ox_diag}. AGN dominated objects lie the top left, star formation dominated objects in the bottom right, which is consistent with the interpretation of \cite{Armus:2007}.
\begin{figure}
\includegraphics[width=8.5cm]{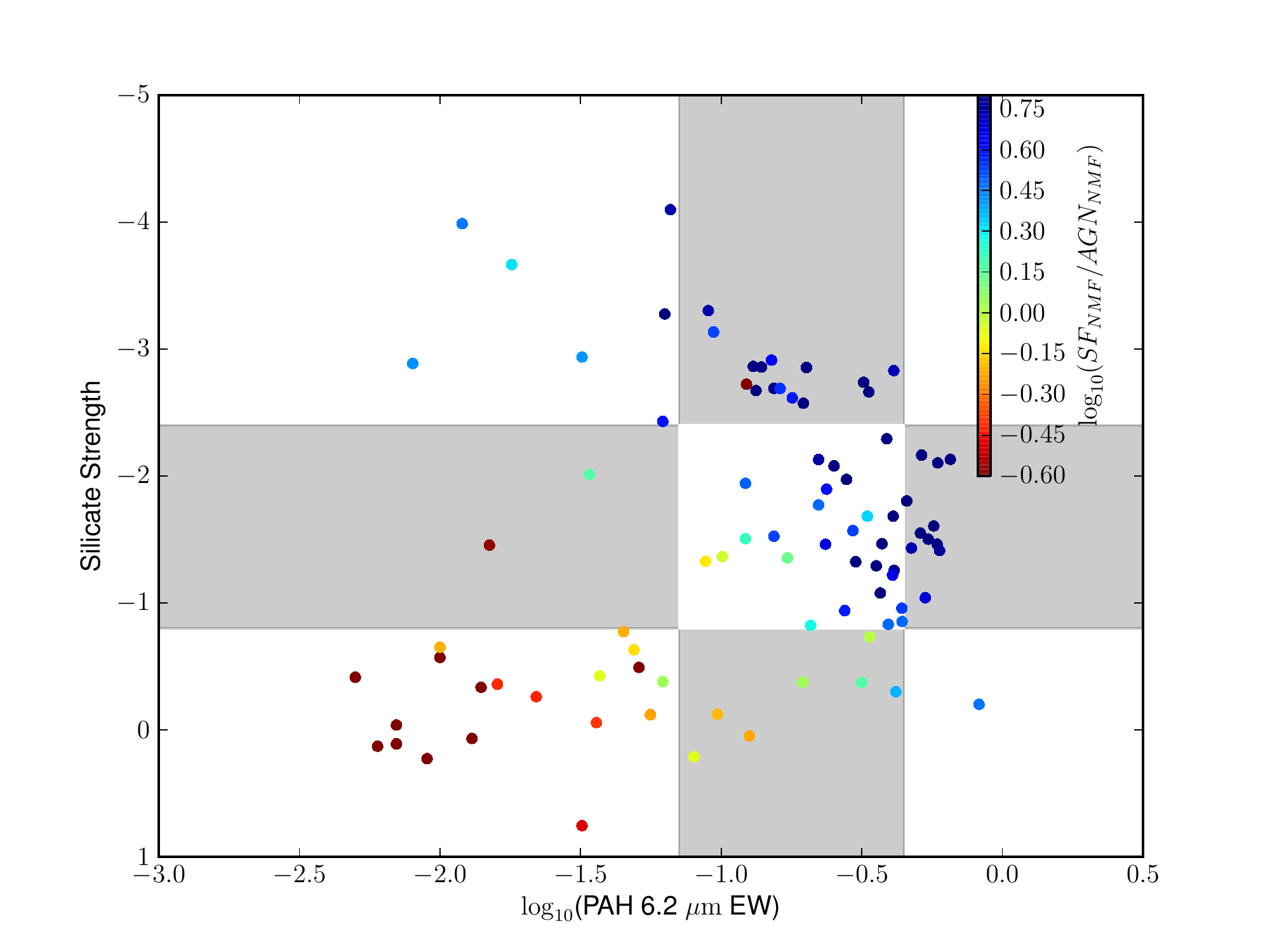}
\caption{The \protect\cite{Spoon:2007uq} diagram showing Silicate strength versus the 6.2 $\mathrm{\mu m}$ PAH equivalent width. The plot is separated into the different Spoon classes and objects are colour coded by the NMF diagnostic.}\label{Spoondiag}
\end{figure}
Our final comparison is with \cite{Spoon:2007uq} diagram, classifying ULIRGs via the strength of their $9.7 \mathrm{\mu m}$ silicate feature and 6.2 $\mathrm{\mu m}$ equivalent width. Figure \ref{Spoondiag} shows 89 ULIRGs in the \cite{Spoon:2007uq} diagram, colour coded by our NMF diagnostic. Our NMF diagnostic suggests AGN dominated objects are on the horizontal branch, while objects on the diagonal branch appear to have significant activity from star formation and AGN. Objects dominated by star formation lie at the extreme right of the two branches. Our diagnostic is consistent with the interpretation of \cite{Spoon:2007uq}.

We have shown that our diagnostic for determining the AGN/star formation ratio is consistent with MIR diagnostic diagrams already in use. Our diagnostic however has the advantage that it uses a far greater wavelength range than current diagnostics and does not rely on specific line measurements. By using 5 of the 7 components in $NMF_{7}$, our diagnostic is also flexible enough to account for the difference in spectra amongst star formation or AGN dominated objects.

%% file: GMM_table_7_1000s_0p1eps_v4.tex
\begin{tabular}{|c|c|c|c|c|c|c|c|}
\hline
\multicolumn{8}{|c|}{GMM and ATLAS classification for 7 templates} \\ \hline\hline
Cluster & prob. & Sy1 & Sy2 & MIR AGN1 & MIR AGN2 & MIR SB & Sbrst \\ \hline
1& 0.301& 90.9& 37.7& 97.5& 52.3&  1.6&  6.2 \\
2& 0.287&  0.0& 17.0&  0.0&  1.7& 43.6& 68.8 \\
3& 0.156&  0.0& 28.3&  0.8& 24.1& 11.7& 12.5 \\
4& 0.147&  9.1& 17.0&  0.8&  8.6& 21.4&  6.2 \\
5& 0.080&  0.0&  0.0&  0.0&  8.0& 20.2&  0.0 \\
6& 0.022&  0.0&  0.0&  0.8&  2.3&  0.8&  6.2 \\
7& 0.004&  0.0&  0.0&  0.0&  1.1&  0.8&  0.0 \\
8& 0.003&  0.0&  0.0&  0.0&  1.7&  0.0&  0.0 \\
\hline
\end{tabular}

%% file: Conclusions_v2.tex
\section{Conclusions}\label{sec:conclusions}
We have carried out the first empirical attempt at learning the fundamental MIR spectral components of galaxies via the multivariate analysis technique, NMF. We have chosen NMF as the most appropriate matrix factorisation technique for our problem as the non negative constraints required by the algorithm, more closely resembles the physical process of emission in the MIR than techniques used in previous studies \citep{LingyuPCA,Hurley:2012}. 
The NMF algorithm has been applied to 729 galaxy spectra, taken from the CASSIS database \citep{Lebouteiller:2011} with spectral redshifts ranging from $(0.01<z<0.2)$. 

We have investigated the number of components needed to accurately reconstruct spectra by evaluating the Bayesian evidence with the nested sampling routine, MULITNEST. The Bayes factor suggests that the number of components exceeds 14 but the gain in increasing the number of components decreases dramatically from seven components onwards. An NMF set with a large number of components may accurately reconstruct all spectra, but assigning physical interpretation to each component becomes difficult, limiting its practical utility.

We have therefore examined the simpler component sets $NMF_{5}$-$NMF_{10}$. We find that despite an increase in the allowed number of components, many of the components remain similar. For example, similar counterparts to components in $NMF_{5}$ can be found in $NMF_{6}$ and above, the sixth component in $NMF_{6}$ can be found in $NMF_{7}$ and above and so on. Finding similar components, despite an increase in flexibility, suggests these components are fundamental spectral components. 

We find the components also have clear, physical interpretation. The first component contains the forbidden fine structure lines associated with narrow line regions and AGN as well as a hot dust continuum also typical of AGN tori. The second common component shows silicate emission at 10 and 18 $\mathrm{\mu m}$ and is indicative of the warm dust associated with both the inner wall of the AGN torus or narrow line region clouds. The third component is a star formation component, containing all of the PAH and molecular hydrogen emission lines, found near PDRs. As the number of components is increased, the colder dust slope is removed to the sixth and seventh components. We interpret this as the separation of unobscured star-forming component (or PDR) from an obscured star-forming component showing colder dust.

Re-running the NMF algorithm on objects dominated by star formation, we show that the PAH emission begins to separate out into two components, which show similar features to the two different PDR components found in \cite{Berne:2007}. 

We have shown that a simpler NMF set with seven components is capable of reproducing the general continuum shape for variety of extragalactic spectra seen in the MIR, though the components struggle with the variation in emission lines. By examining the contributions each component makes to well known objects and previously classified samples, we find different types of objects lie in different regions of 'NMF space'.

Using Gaussian Mixtures modelling, we provide a classification scheme that uses all seven components to separate objects into five different clusters: A Seyfert one cluster, Seyfert two cluster, starburst cluster, dusty and obscured cluster and a type 1.5 Seyfert cluster. Our classification outperforms the Spoon diagram in separating out Seyfert one and two like objects. Unlike the SPoon classification, ours use the whole MIR region, allowing objects without the $9.7 \mathrm{\mu m}$ silicate feature and 6.2 $\mathrm{\mu m}$ equivalent width to be classified. Our GMM based classification can also provide an estimate of the probability of finding a particular galaxy in one of the five clusters.

We also use five of the components to create a star formation/AGN diagnostic which performs well against current MIR diagnostic diagrams. Our NMF based diagnostic has the advantage of considering a greater wavelength range, and can therefore be used for objects where specific emission features have not been observed, or for where spectra are too noisy.

Our NMF components provide fundamental, physical components which are ideal for separating out different types of objects and investigating the power associated with AGN and star formation. They are linked to the actual physical environments such as AGN and star formation unlike templates based on specific objects (e.g. M82) or average templates based on a sample of galaxies. We believe our NMF set could be used to predict useful measures such as star formation rate and AGN luminosity and will investigate this in a future paper. We also believe our NMF set is ideal for more galaxy evolution based investigations such as decomposing the MIR luminosity function into contribution from AGN and and star formation. Our NMF components and code for classification are made available at \url{https://github.com/pdh21/NMF_software/} and on the arxiv.

%% file: ATLAS_temp_table.tex
\onecolumn
\begin{deluxetable}{l c c c c c c l} 
\tabletypesize{\scriptsize}
\tablewidth{0pt}
\tablehead{
\colhead{name} & \colhead{Nsources} & \colhead{$z_{min}$} & \colhead{$\langle z \rangle$} & \colhead{$z_{max}$} & 
\colhead{$\lambda_{min}$ [\uu]} & \colhead{$\lambda_{max}$ [\uu]} & \colhead{comments}   }
\startdata
  Sy1 &  11 &   0.002 &   0.041 &   0.205 &     5.2 &    24.6 &                      Seyfert 1 with $\nu$L$\nu$(7\uu) $<$ 10$^{44}$ erg s$^{-1}$\\
  Sy1x &  72 &   0.003 &   0.091 &   0.371 &     5.0 &    24.6 &                                  intermediate Seyfert types (1.2, 1.5, 1.8, 1.9)\\
  Sy2 &  53 &   0.003 &   0.045 &   1.140 &     5.2 &    24.6 &                      Seyfert 2 with $\nu$L$\nu$(7\uu) $<$ 10$^{44}$ erg s$^{-1}$\\
LINER &  16 &   0.001 &   0.034 &   0.322 &     5.2 &    24.6 &                          LINER with $\nu$L$\nu$(7\uu) $<$ 10$^{44}$ erg s$^{-1}$\\
  QSO & 125 &   0.020 &   1.092 &   3.355 &     2.5 &    24.6 &             QSO1 and Seyfert 1 with $\nu$L$\nu$(7\uu) $>$ 10$^{44}$ erg s$^{-1}$\\
 QSO2 &  65 &   0.031 &   1.062 &   3.700 &     3.6 &    24.6 &             QSO2 and Seyfert 2 with $\nu$L$\nu$(7\uu) $>$ 10$^{44}$ erg s$^{-1}$\\
Sbrst &  16 &   0.001 &   0.091 &   1.316 &     5.2 &    24.6 &               Starburst or HII with $\nu$L$\nu$(7\uu) $<$ 10$^{44}$ erg s$^{-1}$\\
     ULIRG & 184 &   0.018 &   0.730 &   2.704 &     4.5 &    24.6 &                                            ULIRG (low and high redshift sources)\\
       SMG &  51 &   0.557 &   1.869 &   3.350 &     4.8 &    12.0 &                                                           Submillimiter Galaxies\\
  MIR\_AGN1 & 119 &   0.002 &   0.455 &   2.190 &     4.0 &    24.6 &                                          MIR selected AGN with silicate emission\\
  MIR\_AGN2 & 160 &   0.002 &   0.549 &   2.470 &     4.5 &    24.6 &                                        MIR selected AGN with silicate absorption\\
    MIR\_SB & 257 &   0.001 &   0.413 &   2.000 &     4.6 &    24.6 &                                                          MIR selected starbursts\\

\enddata
\caption{Classification of sources by \protect\cite{Hernan-Caballero:2011}}\label{tabla_templates}
\end{deluxetable}